\documentclass[12pt]{article}
\usepackage{amsfonts,amssymb,amsbsy,amsmath,amsthm,latexsym,natbib,graphicx}
\usepackage{epsf}
\usepackage[english]{babel}
\usepackage[utf8]{inputenc}
\usepackage{algorithm}
\usepackage[noend]{algpseudocode}
\usepackage{subfigure,color,comment}
\usepackage{hyperref}

\newtheorem{theorem}{Theorem}
\newtheorem{proposition}{Proposition}
\newtheorem{lemma}{Lemma}

\theoremstyle{definition}

\newtheorem{remark}{Remark}

\newcommand{\mubar}{\overline{\mu}}

\newcommand{\pbar}{\overline{p}}
\newcommand{\cbar}{\overline{c}}

\newcommand{\sigmasqbar}{\overline{v}}

\newcommand{\Prob}[1]{\mathbb{P} \left({#1}\right)}

\newcommand{\Expect}[1]{\mathbb{E} \left[{#1}\right]}

\newcommand{\md}{\mbox{d}}

\newcommand{\calW}{\mathcal{W}}
\newcommand{\calO}{\mathcal{O}}

\graphicspath{{./}}
\title{Fast return-level estimates for flood insurance via an improved Bennett inequality for random variables with differing upper bounds}
\author{Anna Maria Barlow$^1$\footnote{anna-maria.aulbach@mathematik.uni-stuttgart.de} and Chris Sherlock$^2$\footnote{c.sherlock@lancaster.ac.uk}}
\date{{\small $1$ Institute for Stochastics and Applications, Universitat Stuttgart, Germany.}\\{\small $^2$Department of Mathematics and Statistics, Lancaster University, UK.}}

\begin{document}
\maketitle
%\begin{center}
%\textbf{Abstract}
%\end{center}
\begin{abstract}
   Insurance losses due to flooding can be estimated by simulating and then summing losses over a large number of locations and a large set of hypothetical years of flood events. Replicated realisations lead to Monte Carlo return-level estimates and associated uncertainty. The procedure, however, is highly computationally intensive. We develop and use a new, Bennett-like concentration inequality to provide conservative but relatively accurate estimates of return levels. Bennett's inequality accounts for the different variances of each of the variables in a sum but uses a uniform upper bound on their support. Motivated by the variability in the total insured value of risks within a portfolio, we incorporate both individual upper bounds and variances and obtain tractable concentration bounds. Simulation studies and application to a representative portfolio demonstrate a substantial tightening compared with Bennett's bound. We then develop an importance-sampling procedure that repeatedly samples annual losses from the distributions implied by each year's concentration inequality, leading to conservative estimates of the return levels and their uncertainty using orders of magnitude less computation. This enables a simulation study of the sensitivity of the predictions to perturbations in quantities that are usually assumed fixed and known but, in truth, are not. 
\end{abstract}

\section{Introduction}
\label{sec.intro}
The UK winter of 2015--16, which included Storms Desmond and Eva, saw more than 20,000 businesses and residential properties flood, resulting in insurance claims totalling nearly \pounds 480 million. The floods of 2013--14 saw similar claims, and the floods in the summer of 2007 were even more costly \citep{GOV}.

We are motivated by the problem of estimating the future insurance costs for a portfolio over the course of many hypothetical years, each of which may contain several flood events. The UK government's 2015 solvency regulation \cite[]{Swain2015} requires specific information on the 200-year return level, the level of insurance payout over a year that is expected to be exceeded once in every 200 years on average. Other return levels from the 2-year to the 500-year and above are also of interest to insurers.

Each portfolio consists of a number of insurance \emph{risks}, and each risk, $r$, is partitioned into a number, $n_r^{(\mathrm{sub})}$, of \emph{subrisks}; the value, $b_r$, of the risk is partitioned equally among the subrisks, a common practice in the industry.  The procedure for estimating the distribution of each return level that is used by a funder of this work, for example, involves simulating both the number of flood events in a year ($n_y^{(\mathrm{ev})}$ for year $y$) and the flood events themselves from a statistical extreme-value model \cite[][]{Lambetal2010,KeefTawnLamb2013,ToweTawnLambSherlock2019} based on the conditional multivariate extremes model of \cite{HeffernanTawn2004}. The extremes model targets measurements across a large network of river gauges; see Appendix \ref{sec.EVTmodel} for more details about the flood-event model. This produces information on a large number, $n_y$, of years of flood events; typically $n_y=1000$ or $n_y=10000$. Our motivating insurance portfolio and associated flood characteristics for each year and event are summarised in Section \ref{sec.stdPortfolio}.

Our contribution concerns the simulation-based methodology that \emph{uses} the output from the extreme-value model for the $n_y$ years to estimate the $k$-year return levels of insurance losses and provide a measure of uncertainty in these estimates. Inevitably there is some associated jargon and notation, which we bring together for reference in Table \ref{tab:Defs}.

\subsection{Estimating return levels}
\label{subsec.estimate.return.level}
For each year, $y$, each flood event, $e$, simulated from the multivariate extremes model is processed. For each risk, $r$, this provides a probability of flooding, $p_{y,e,r}$, and values $\alpha_{y,e,r}$ and $\beta_{y,e,r}$ defining the $\mathsf{Beta}(\alpha_{y,e,r},\beta_{y,e,r})$ distribution for the \emph{damage ratio}, the fraction of the insured value that will be claimed should the property flood; together these are called the \emph{loss distribution}. For each subrisk within that risk, both whether there is a claim or not and, if there is a claim, the fraction of the insured value claimed are simulated independently using the loss distribution associated with the parent risk. See \cite{Oasis2020} for the overarching simulation-based approach and \cite{Lambetal2010} for a similar methodology applied to flood events. Importantly, the flood-event model is assumed to account for all of the dependence between losses at different subrisks; the individual losses are then assumed to be independent conditional on the individual loss distributions that arise from the model. In a particular year we are interested in the \emph{total loss} for the portfolio over the year: 
\begin{equation}
  \label{eq.AddItAll}
  T_y=\sum_{e=1}^{n^{(\mathrm{ev})}_y}\sum_{r\in\mathcal{R}_e}\frac{b_r}{n^{(\mathrm{sub})}_r}\sum_{s=1}^{n_r^{(\mathrm{sub})}}Z_{y,e,r}^{(s)},
  ~~~y\in\{1,\dots,n_y\}.
\end{equation}

Here $\mathcal{R}_e$ is the set of risks with non-zero probability of flooding during event $e$ of year $y$  
and $Z_{y,e,r}^{(s)}$ is the fractional loss from subrisk $s$ of risk $r$ during event $e$ of year $y$: $Z_{y,e,r}^{(s)}=0$ with a probability of $1-p_{y,e,r}$ and otherwise $Z_{y,e,r}^{(s)}\sim \mathsf{Beta}(\alpha_{y,e,r},\beta_{y,e,r})$. \footnote{For simplicity here we ignore the possibility that a subrisk subject to two floods in a year might have an altered loss distribution for the second flood, although it is possible to deal with this.}. 

Simulating each of the $n$ losses once provides a single Monte Carlo estimate of the loss for a given year. This is repeated for each of the $n_y$ years of events from the multivariate extremes model. For each $k$ of interest ($k\in\{2,5,10,20,50,100,200,500\}$ for us), the $(1-1/k)$th quantile of the yearly totals provides a single Monte Carlo estimate of the $k$-year return level. Repeating this procedure a large number, $M$, of times (typically $M=100$ or $M=1000$) provides a distribution for each $k$-year return level, from which are obtained a mean point estimate and a (quantile-based) 95\% prediction interval.

The standard method is computationally costly in real scenarios where the portfolio size might be between $10^5$ and $5\times 10^6$ as it involves the simulation of multiple replicates of an exceedingly large number of random variables, even accounting for the fact that we can automatically set $Z_{y,e,r}=0$ for risks with no chance of flooding during a particular event. Let $\overline{n}^{(\mathrm{ev})}$, $\overline{n}^{(p>0)}$ and $\overline{n}^{(\mathrm{sub})}$ be, respectively, the mean number of: events per year, risks with non-zero probability of flooding averaged over events and subrisks per risk. Furthermore, define the average probability of flooding amongst those risks that might flood: $\pbar:=\sum p_{y,e,r}/\sum 1(p_{y,e,r}>0)$, where each sum is over all $n_y$ years, all $n_y^{\mathrm{ev}}$ events in that year and all $|\mathcal{R}_e|$ risks for that event. Then the number of simulations is $\mathcal{O}(Mn_y\overline{n}^{(\mathrm{ev})}\overline{n}^{(p>0)}\pbar \overline{n}^{(\mathrm{sub})})$. In practice, the Monte Carlo procedure can take 24 hours or more to complete for a given portfolio. 

This article presents a much faster, yet relatively accurate and provably conservative alternative methodology for estimating the $k$-year return levels for various $k$. It uses probabilistic bounds on each yearly total loss based on an improvement to \emph{Bennett's} inequality. For our large portfolio, with $M=1000$ simulations using $n_y=1000$ years, this reduces the computational time from around $7.5$ CPU hours to $2.9$ CPU seconds. This massive reduction permits exploration of alternative scenarios which we demonstrate by conducting a short study of sensitivity to mis-specification of the expected damage ratio.

\subsection{Bennett's and Hoeffding's Inequalities}

 Let $Z_{y,e,r}^{(j)}$ correspond to the $j$th realisation from a sequence of independent random variables each of which is $0$ with a probability of $1-p_{y,e,r}$ and is otherwise drawn from the beta variable unique to that $(y,e,r)$ combination.

Centering the individual losses leads to $X_i=Z_i-\Expect{Z_i}$ ($i=1,\dots,n$), and $T_n=\sum_{i=1}^nX_i+\sum_{i=1}^n \Expect{Z_i}$. Before introducing concentration inequalities, we first formalise the set up that will be used throughout this article. Let $X_1,\dots,X_n$ be independent random variables with
\begin{equation}
  \label{eq.X.moments}
  \Expect{X_i}=0
  ~~~\mbox{and}~~~
  \mathsf{Var}[X_i]=\sigma^2_i,
\end{equation}
and  define
\begin{equation}
  \label{eq.define.S.sigmasqbar}
  S_n=\sum_{i=1}^n X_i
  ~~~\mbox{and}~~~
  \sigmasqbar=\frac{1}{n}\sum_{i=1}^n \sigma_i^2.  
\end{equation}

Losses due to flooding typically possess the following two properties:
\begin{enumerate}
\item
  The total value insured for a given risk, and hence the maximum loss from that risk, varies from risk to risk, often by more than an order of magnitude.
  \item
  In a given event, for most risks (but not all) the probability of flooding is small, and even conditional on flooding, the total damage is often much less than the value insured; hence the variance for that risk is smaller than might be guessed given the value insured.
\end{enumerate}

The first property suggests that Hoeffding's inequality might be appropriate, since it accounts for the individual variations in support:

\begin{theorem}\label{thrm.Hoeff}\cite[][Theorem 2]{Hoeffding1963} Let $X_1,\dots,X_n$ and $S_n$ be as defined in and above equations \eqref{eq.X.moments} and \eqref{eq.define.S.sigmasqbar} and let $\Prob{a_i \le X_i\le c_i}=1$ for some $a_1,\dots, a_n$ and $c_1,\dots,c_n$. Then 
  \[
  \frac{1}{n}\log \Prob{S_n\ge nt}\le -\frac{2t^2}{\frac{1}{n}\sum_{i=1}^n(c_i-a_i)^2}.
  \]
\end{theorem}

The second property suggests the use of Bennett's inequality, which, unlike Hoeffding's inequality takes the variances into account and allows for different variances for each risk, but does not account for the variations in support, assuming an upper bound, $c_* > 0$, common to all risks: for each $i$ in $1,\dots,n$,
\begin{equation}
  \label{eq.uniform.support}
  \Prob{X_i\le c_*}=1.
\end{equation}

\begin{theorem}\cite[]{Bennett1962}
  \label{theorem.Bennett}
  Let $X_1,\dots,X_n$ be independent random variables subject to \eqref{eq.X.moments} and \eqref{eq.uniform.support}. Then for any $t>0$,
  \begin{equation}
    \label{eq.Bennett}
    \frac{1}{n}\log     \Prob{S_n>nt}
    \le
    -\frac{\sigmasqbar}{c_*^2}h\left(\frac{c_* t}{\sigmasqbar}\right),
    \end{equation}
  where $h(u)=(1+u)\log(1+u)-u$ and $S_n$ and $\sigmasqbar$ are as defined in \eqref{eq.define.S.sigmasqbar}.
\end{theorem}

Neither inequality is entirely satisfactory as each fails to account for an important property of the components of the sum that makes up the total loss in a year.

Since its inception, Bennett's inequality has been tightened slightly \cite[Theorem 3]{Zheng2018,Hoeffding1963} subject to the same conditions as the original theorem, and tightened further subject to additional moment information on the individual $X_i$ \cite[e.g.,][]{PinelisUtev1990,Light2021concentration,BentkusJuskevicius2008,Pinelis2014}. All of the listed articles, however, require a uniform upper bound \eqref{eq.uniform.support}; for an exception, see \cite{Bennett1962} Equation 2c, mentioned in Section \ref{sec.discussion} of this article.

%Zheng (2018) CITE provides two direct improvements on the basic Bennett inequality, one of which uses a lower bound on the gap between the arithmetic mean of a set of positive numbers the corresponding geometric mean. 

%The current procedure involves simulating $m=100$ replicates of the yearly losses from each of $10^4$ years worth of simulated events. Portfolio sizes vary by orders of magnitude but can be up to $10^7$. Even though for each event no simulations are needed for risks with zero probability of flooding during that event, the simulations of losses typically take days to complete (CITE source? see ANNA'S THESIS? for more details). CITE Anna's Thesis (unpublished) uses conservative bounds on the sums of losses to reduce the simulation cost. This article derives tighter bounds that is suitable for these purposes.

Theorem 1.3 of \cite{Bentkus2004} provides a Hoeffding-like bound that only requires an upper bound on the support, but in cases of interest to us is looser than the bound of Hoeffding. \cite{Light2021concentration}  uses moment information from the individual $X_i$ to improve upon both Hoeffding's inequality and Bennett's inequality. The Hoeffding improvement is potentially useful when the $X_i$ are identically distributed or when there are a few fixed $t$ values of interest. When the variables are not identically distributed, evaluating $\Prob{S_n>nt}$ requires a fresh $\mathcal{O}(n)$ calculation for each new $t$. When simulating via the inverse-cumulative distribution function (inverse-cdf) method, as we will do, we must find the minimum $t$ such that $\Prob{S_n \ge t}\le \epsilon$, numerical solution of the equation would require many $\mathcal{O}(n)$ calculations, which we wish to avoid.

We focus on Bennett's inequality and tighten it by accounting for the different supports of the individual subrisks.  

\subsection{Our contribution}

Given a bound on the probability of the loss for a particular year exceeding a given amount, our first methodological contribution is to avoid simulating the individual losses at each individual subrisk for each event for that year and summing them; instead, we simulate from the distribution implied by the bound on the sum for that year. Repeating for each year leads to conservative estimates and upper prediction limits for the $k$-year return levels for all required $k$. This procedure can be repeated using equivalent concentration bounds on the lower tails of the yearly losses. Whilst typically faster than direct Monte Carlo simulation, the method does not give the desired improvement in computational effort. Our second methodological contribution is an importance-sampling resampling scheme that speeds up the procedure by several orders of magnitude compared with the standard methodology. Our scientific contribution is a sensitivity study enabled by this speed up. We show that in our portfolio, failing to account for even a moderate amount of uncertainty (known to exist in practice) in the expected damage ratio at each risk makes little difference to the estimated return-level quantiles.  

Section \ref{sec.theory} of this article describes the new Bennett-like inequalities which allow for different supports for the individual components of the sum. In Section \ref{sec.numerical}, the new bounds are compared with existing bounds and estimates in various binary-data scenarios and in four indicative years from a realistic, simulated portfolio and set of flood events. The new scheme, including the importance-sampling resampling method is described in Section \ref{sec.estimate.return.level}, and timings and estimates of the $k$-year return levels are compared with those from the standard scheme using the full portfolio. The section then describes the sensitivity study, and the article concludes in Section \ref{sec.discussion} with a discussion. %The section then describes a sensitivity study on the impact of mis-specification of the damage-ratio. The article concludes with a discussion.

\section{Bennett inequalities allowing for different support}
\label{sec.theory}

Suppose that we have information on the support of each variable, replacing \eqref{eq.uniform.support} with
\begin{equation}
  \label{eq.individual.support}
  \Prob{X_i\le c_i}=1, ~~~i=1,\dots,n.
\end{equation}
Let $Z_i$ be the loss from an individual risk during a large event, such as a flood. Then $0\le Z_i\le b_i$, where $b_i$ is the maximum value insured. Defining $X_i=Z_i-\mathbb{E}[Z_i]$ we see that \eqref{eq.individual.support} holds with $c_i=b_i-\mathbb{E}[Z_i]$. Clearly, \eqref{eq.Bennett} continues to hold in this case, with $c_*=\max_{i=1,\dots,n}c_i$; however, as discussed in Section \ref{sec.intro}, and shown in the numerical experiments of Section \ref{sec.numerical}, Bennett's inequality using $c_*$ is unsatisfactory.

Fundamental to the proofs of Theorems \ref{thrm.Hoeff} and \ref{theorem.Bennett} and to many of the improvements upon them is the Chernhoff bound \cite[e.g.,][]{BoucheronLugosiMassart2013}: for any $\lambda>0$,
\[
\Prob{S_n\ge nt} =\Prob{\exp(\lambda S_n)\ge \exp(\lambda nt)}
\le \exp(-\lambda nt)\Expect{\exp(\lambda S_n)},
\]
by Markov's inequality. Thus
\begin{equation}
  \label{eq.Chernhoff}
  \frac{1}{n}\log \Prob{S_n\ge nt}
  \le
  \inf_{\lambda >0} \left\{\frac{1}{n}\log\Expect{\exp(\lambda S_n)} -\lambda t\right\}.
\end{equation}

Theorem \ref{theorem.Bennett} uses the following bound on the moment generating function of $S_n$ in \eqref{eq.Chernhoff}:
\[
\frac{1}{n}\log \Expect{\exp(\lambda S_n)}\le\lambda^2\sigmasqbar f_2(\lambda c_*),
\]
where, for $k=1,2,\dots$,
\[
f_k(u)
=
\frac{1}{u^k}\left\{\exp(u)-\sum_{j=0}^{k-1} \frac{u^j}{j!}\right\}
=
\sum_{j=0}^\infty \frac{u^{j}}{(j+k)!}.
\]
Our main result, Theorem \ref{thrm.new}, relies on the following improved bound; both are proved in Appendix \ref{sec.proofs}.

\begin{lemma}
  \label{lem.mgf.bound}
  \begin{align}
    \label{eq.new.mgf.bound.tightest}
    \frac{1}{n}\log \Expect{\exp(\lambda S_n)}
    &\le
    \frac{1}{2}\lambda^2 \sigmasqbar+\lambda^2 K\left\{f_2(\lambda c_*)-\frac{1}{2}\right\}
-
\lambda^4c_*^2 K_1 f_4(\lambda c_*)\\
\nonumber
&=:    B(\lambda; c_{*},\sigmasqbar,K,K_1),
  \end{align}
where, 
\begin{equation}
  \label{eq.define.Ks}
K=\frac{1}{n}\sum_{i=1}^n \sigma^2_i\frac{c_i}{c_*}
~~~\mbox{and}~~~
K_1=\frac{1}{n}\sum_{i=1}^n \sigma_i^2\frac{c_i}{c_*}\left(1-\frac{c_i}{c_*}\right).
\end{equation}
  Furthermore, a lower bound on the upper bound on $\frac{1}{n}\log \Expect{\exp(\lambda S_n)}$ that could be achieved if the methodology which generates Theorem \ref{theorem.Bennett} could take \eqref{eq.individual.support} into account exactly, is
  \begin{equation}
    \label{eq.idealised}
B_{\mathrm{lb}}:=B(\lambda; c_{*},\sigmasqbar,K,K_1)-\lambda^5c_*^3(K-K_1)f_5(\lambda c_*).
  \end{equation}
  \end{lemma}

Bound \eqref{eq.new.mgf.bound.tightest} relies only on the summary statistics $c_{*}$, $\sigmasqbar$, $K$ and $K_1$, and the resulting minimisation problem can be efficiently solved numerically. Alternatively, the looser bound obtained by omitting the term in $K_1$ leads to a tractable minimisation problem with a solution in terms of Lambert's W function \cite[e.g.,][]{Corless1996}, which we denote by $\calW$; the solution can then be substituted into \eqref{eq.new.mgf.bound.tightest} or into the looser bound. These considerations lead to our main result:
  \begin{theorem}
    \label{thrm.new}
  Let $X_1,\dots,X_n$ be independent random variables subject to \eqref{eq.X.moments} and \eqref{eq.individual.support}. Define $S_n$, $\sigmasqbar$, $B$, $K$ and $K_1$ as in \eqref{eq.define.S.sigmasqbar}, \eqref{eq.new.mgf.bound.tightest} and \eqref{eq.define.Ks}, and let $c_*=\max_{i=1,\dots,n}c_i$. For any $t>0$,
    \begin{align}
      \label{eq.tightestP}
\frac{1}{n}\log    \Prob{S_n\ge nt}
&    \le B_1(t):=
    \inf_{\lambda>0}\{B(\lambda; c_{*},\sigmasqbar,K,K_1)-\lambda t\}\\
    \label{eq.laxerP}
&    \le B_2(t):=
    B(\lambda_*; c_{*},\sigmasqbar,K,K_1)-\lambda_*t\\
        \label{eq.laxestP}
& \le B_3(t):=
    B(\lambda_*; c_{*},\sigmasqbar,K,0)-\lambda_* t\\
\label{eq.looser.problem}
&=    \inf_{\lambda>0}\{B(\lambda; c_{*},\sigmasqbar,K,0)-\lambda t\},
    \end{align}
    where 
    \begin{equation}
      \label{eq.lamstar}
      \lambda_*=
      \frac{1}{c_*}\left[
        \frac{K+tc_*}{\sigmasqbar -K}
        - \calW\left(\frac{K}{\sigmasqbar-K}
        \exp\left(        \frac{K+tc_*}{\sigmasqbar -K}
\right)\right)
        \right].
    \end{equation}
  \end{theorem}

If it is known that $c_i\ge c_0>0$ for all $i>0$ then a tighter bound may be obtained from an analog of Lemma \ref{lem.mgf.bound} by linearly interpolating between $\lambda c_0$ and $\lambda c_*$ rather than between $0$ and $\lambda c_*$. Even removing the $K_1$ term, this does not lead to a tractable minimisation problem, but the bound may still be efficiently minimised numerically with respect to $\lambda$. In scenarios of interest to us, $c_0\ll c_*$ and the gains are negligible.

A stronger tightening of \eqref{eq.new.mgf.bound.tightest} is possible at the expense of evaluating summary statistics involving $j$th powers of the $c_i$ and $c_*$: $K_j=(1/n)\sum_{i=1}^n \sigma^2_i (c_i/c_*)(1-c_i^j/c_*^j)$. A brief derivation of this sequence of bounds is given in Appendix \ref{app.tighter.still}. The new bounds can only be substantially tighter if the $K_j$, $j\ge 2$, are substantially larger than $K_1$. Across toy scenarios (i) to (iv) from Section \ref{sec.numerical.toy}, $K_2/K_1$ varied between $1.23$ and $1.36$, and $K_3/K_2$ varied between $1.06$ and $1.11$, showing quickly diminishing returns, which is unsurprising since our entire motivation was that for many subrisks, $c_i\ll c_*$. In other scenarios, the bound on further improvement, obtained through \eqref{eq.idealised}, could inform a decision on gathering further summaries.

  \section{Numerical experiments}
  \label{sec.numerical}
  This section compares the new inequalities with existing ones. We start with four toy examples that explore a variety of regimes. We then introduce our portfolios and explore the performances of the inequalities across four different years that represent the spread of possibilities within our standard portfolio.
  
  \subsection{Toy examples}
  \label{sec.numerical.toy}
We construct four toy scenarios for comparing our bounds by, in each case, simulating $n=10^5$ random distributions as follows. For $i=1,\dots,n$, simulate $b_i$ from some non-negative distribution, $G$, and simulate $p_i\sim \mathsf{Beta}(\alpha,\beta)$ for some $\alpha>0$ and $\beta>0$. Here, $p_i$ represents the probability of flooding and $b_i$ represents the (fixed, for simplicity) loss if a flood occurs; hence, the loss from the $i$th property or risk is $Z_i=b_i\times\mathsf{Bernoulli}(p_i)$. From $Z_i$ we create the centred random variable $X_i=Z_i-b_i p_i$, and we are interested in $S_n=\sum_{i=1}^n X_i$. The individual upper bounds and variances are, respectively, $c_i=b_i(1-p_i)$ and $\sigma^2_i=b_i^2p_i(1-p_i)$.

Our motivation arises from cases where most of the random variables are likely to take low values, so for the first three experiments we take $\alpha=1$ and $\beta=10$. Scenarios of interest to us also have fewer larger $c_i$ than there are smaller $c_i$ and we consider three tail scenarios: (i) light tails, where $B_i=|\zeta_i|$, with $\zeta_i\sim \mathsf{N}(0,1)$; (ii) exponential tails, where $B_i\sim \mathsf{Exp}(1)$ and (iii) heavy tails, where $B_i\sim \mathsf{Pareto}$ with a shape parameter of $\alpha=4$. Our final scenario (iv) uses the exponential tail behaviour from (ii) and has $P_i\sim \mathsf{Unif}(0,1)$. We vary $t$ from $0$ to such a value that our tightest bound gives $n^{-1}\log \Prob{S_n> nt}\approx -0.025$;  even with $n$ as low as $600$, this covers events down to a probability of less than $10^{-6}$.

Theorem \ref{thrm.new} does not require that the random variables $X_i$ be bounded below; however, when they are bounded below by $a_i\le 0$, Hoeffding's inequality (Theorem \ref{thrm.Hoeff}) provides an alternative bound which takes account of the individual ranges, but not the individual variances. In our experiments, we also evaluate the Hoeffding bound for comparison, as well as the tightening of Bennett's bound in Theorem 3 of \cite{Hoeffding1963}, which is the tightest possible under the conditions of Theorem \ref{theorem.Bennett}. The latter is not shown in the plots as in all scenarios it was visually indistinguishable from Bennett's bound.

\begin{figure}
\begin{center}
  \includegraphics[scale=0.44,angle=0]{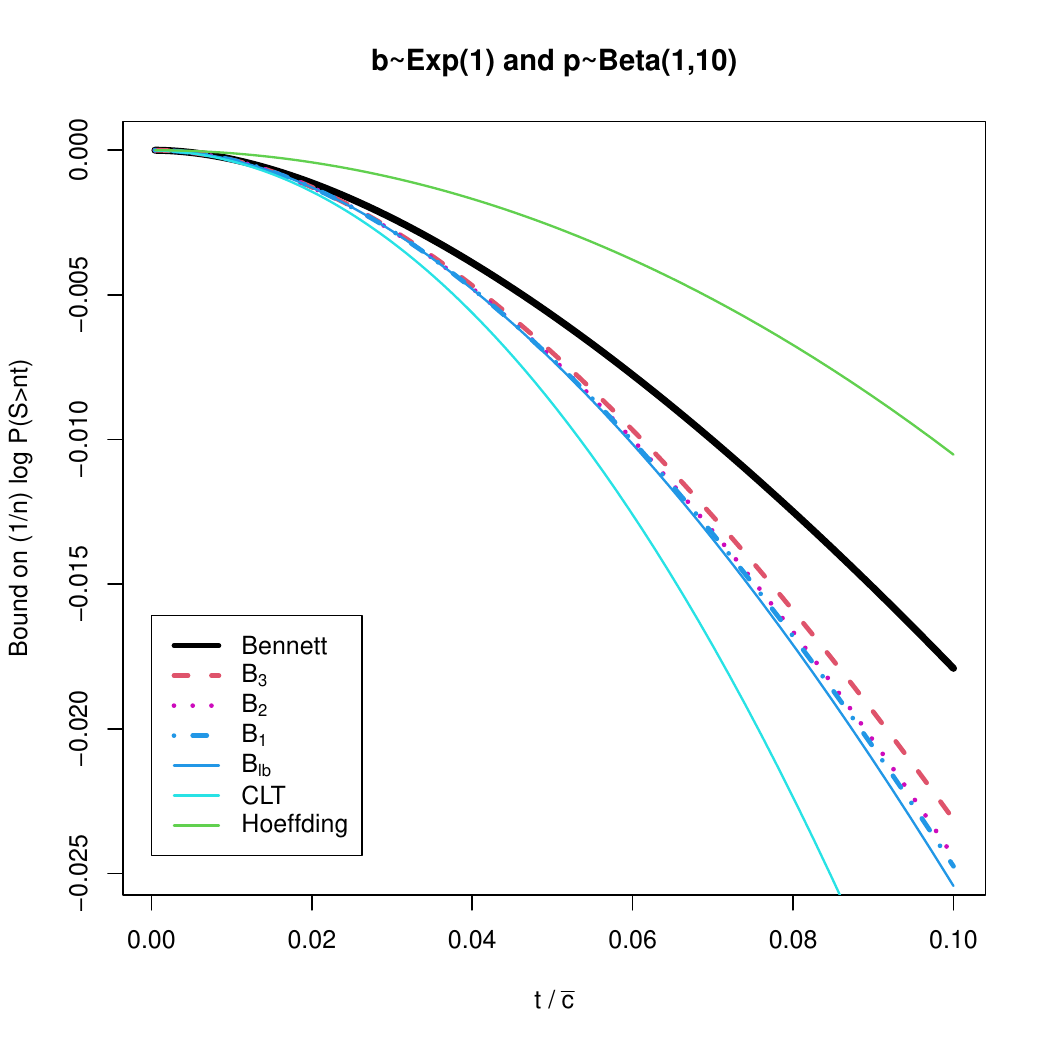}
  \includegraphics[scale=0.44,angle=0]{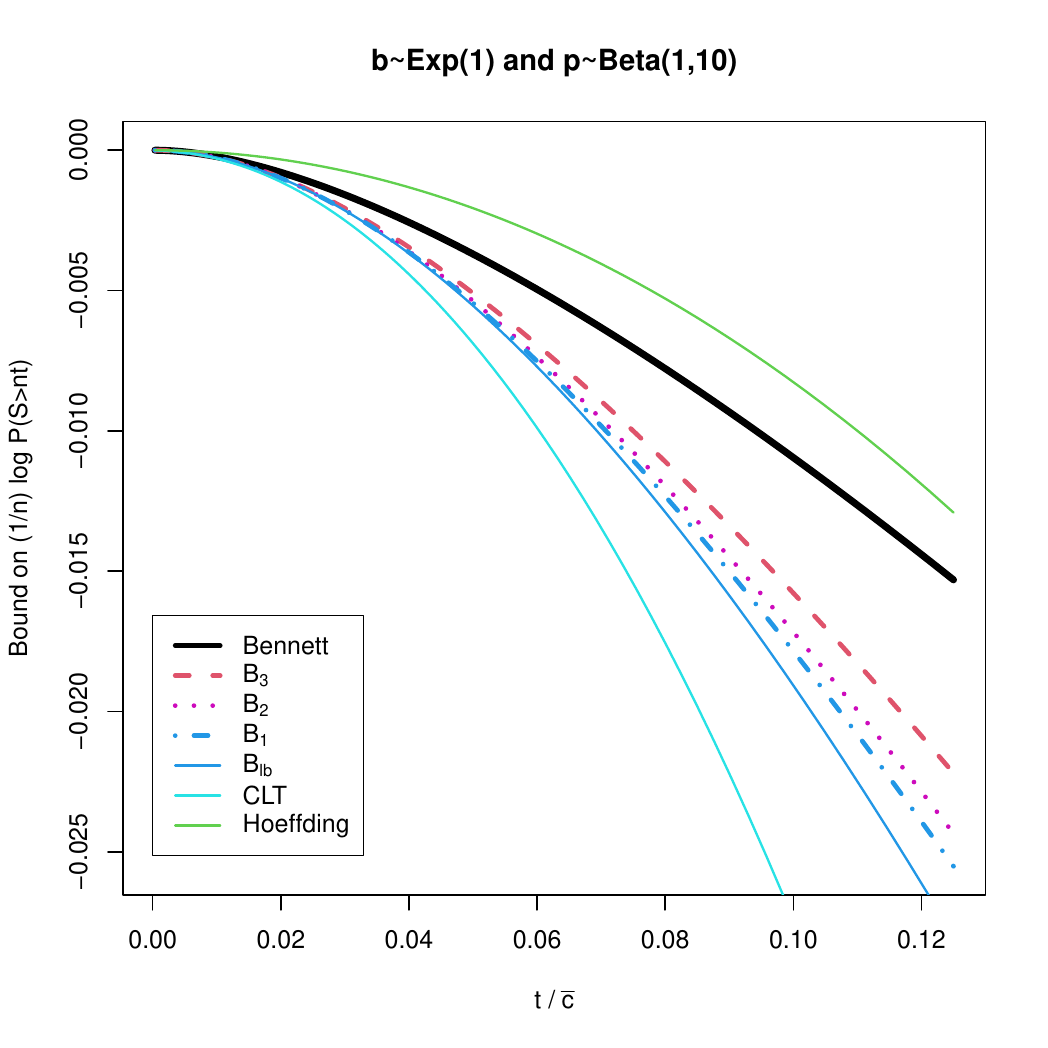}
  \includegraphics[scale=0.44,angle=0]{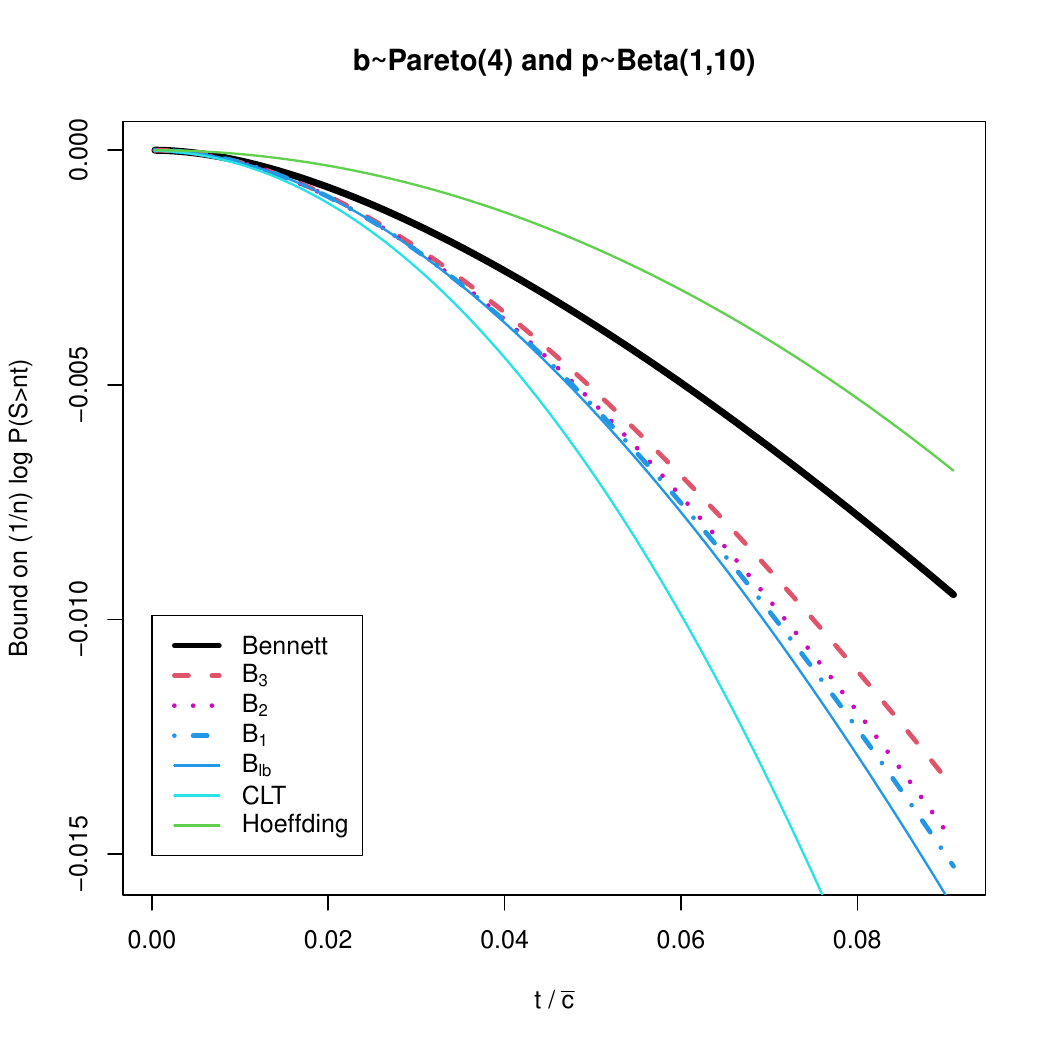}
  \includegraphics[scale=0.44,angle=0]{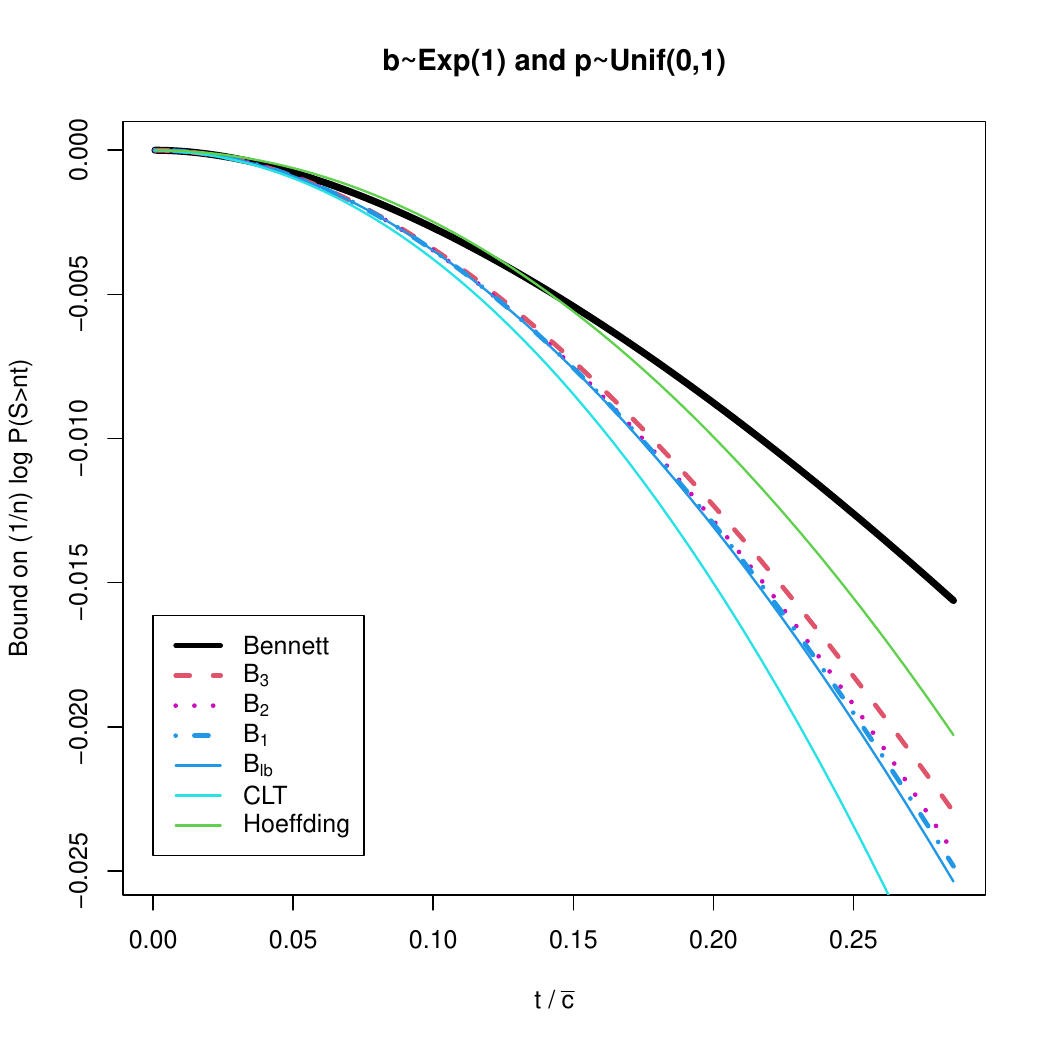}
  \caption{Comparison of bounds on $n^{-1}\log \Prob{S_n\ge nt}$ plotted against $t/\cbar$: Bennett (black solid); $B_1$ \eqref{eq.tightestP} in blue dot-dash; $B_2$ \eqref{eq.laxerP} in magenta dotted and $B_3$ \eqref{eq.laxestP} in red dashed. Other solid lines are: the large-$n$ approximation from the central limit theorem (cyan), the Hoeffding bound (green) and a lower bound, $B_{\mathrm{lb}}$, on the best possible ``Bennett''-like bound, obtained  from \eqref{eq.idealised} (blue).  The four panels correspond to the scenarios as follows: (i) top left, (ii) top right, (iii) bottom left, and (iv) bottom right. The $x$-axes of the panels are chosen so that the $y$-axes display the same range of probabilities.
\label{fig.numerical.expts}
}
\end{center}
\end{figure}

Figure \ref{fig.numerical.expts} shows the bounds plotted against $t$, and includes the corresponding large-$n$ approximation from the central limit theorem, $n^{-1}\log \Prob{S_n\ge nt}\sim -0.5 t^2/\sigmasqbar$, which, for large $n$, may be considered an upper bound on what might be achievable by any concentration inequality. We also include bounds derived from $B_{\mathrm{lb}}$, which bounds the potential improvement from using an idealised Bennett-like inequality that would cost $\mathcal{O}(n)$ to evaluate for each $\lambda$. All of the new bounds improve on the Bennett bound, accounting for a substantial fraction of the gap between the Bennett bound and the approximation from the central limit theorem. Furthermore, the possibility for further improvement by approaching a ``perfect'' Bennett-like bound is small. All of the new bounds are also substantially tighter than the Hoeffding bound over the range of $t$ values considered. Using the $\lambda_*$ obtained from minimising \eqref{eq.laxestP} within the tighter bound \eqref{eq.laxerP} appears to offer a non-negligible improvement over \eqref{eq.laxestP}.

\subsection{Our portfolios}
\label{sec.stdPortfolio}
Our simulated `standard portfolio' is itself based upon a smaller portfolio provided by JBA Risk Management, which consisted of 10,000 risks and 181,789 subrisks. For anonymisation purposes, this was transformed and then perturbed to become our `small portfolio'. Furthermore, 1000 (hypothetical) years of output from the flood-event model supplied, for each flood event and each risk, the probability of flooding and the parameters of the beta distribution of the fraction of the total-insured value that would be claimed were flooding to occur. Recall that each risk is partitioned into subrisks, with the total insured value split evenly across the subrisks (this is a common simplifying assumption made in the industry) and with the loss from each subrisk treated independently of the losses from all other subrisks. 

The portfolio provided by JBA was much smaller than typical portfolios so from it we bootstrapped a portfolio with exactly ten times the total numbers of risks and subrisks; we refer to this as the standard portfolio. If in the original portfolio, subrisk S was part of risk R then the simulation ensured that in the standard portfolio each copy of subrisk S would be part of a copy of risk R. In the standard portfolio, the lowest insured value for a subrisk was \pounds 15,559.14 and the highest was \pounds 379,382.60. In an analogous manner, we also created a `large portfolio' which has ten times the number of risks as our standard portfolio. 

Table \ref{table.four.years} provides summary statistics for four particular years from our standard portfolio:
the years with the lower quartile (labelled A), median (B), upper quartile (C) and maximum (D) expected loss. Across these four years, the mean number of non-zero terms summed varies from $132$ (year A) to $12907$ (year D).

\begin{table}
\caption{\label{table.four.years} Statistics for the years with the lower quartile (A), median (B), upper quartile (C) and maximum (D) expected loss: number of events in that year, $n^{(\mathrm{ev})}$, expected loss ($\Expect{S_n}$), number of (subrisk,event) combinations in that year with a non-zero probability of flooding ($n^{(p>0)}$), mean probability of flooding amongst this (subrisks,event) set ($\pbar$). To understand the ``average'' effect of the ratio $\alpha_{y,e,r}/\{\alpha_{y,e,r}+\beta_{y,e,r}\}$, we also specify the mean over this (subrisk,event) set of the expected fraction of the insured value that would be lost were the subrisk to flood ($\mubar$). Subrisks with a non-zero probability of flooding more than once during the year are treated as separate subrisks for the purpose of these statistics.}
  \begin{center}
\begin{tabular}{c|r|r|r|r|r}
  Year & $n^{(\mathrm{ev})}$&$\mathbb{E}[S_n]$ (\pounds k)& $n^{(p>0)}$& $\pbar$ & $\mubar~(\%)$\\
  \hline
  A&7&739 & 3 220   & 0.041& 9.92\\
%  B&3 922& 5 210  & 0.094& 7.27\\
  B&12&3 923& 2 480  & 0.127& 7.86\\
  C&10&13 740& 8 690 & 0.168& 9.06\\
  D&11&266 800&64 860& 0.199&10.72
\end{tabular}
% number of events with non-zero probability of flooding are 5,4,5,2 resp.
% - one of the two for D was absolutely massive with 366 risks (6738 subrisks) in small portfolio affected.)
\end{center}
\end{table}

\subsection{Flood example}
\begin{figure}
\begin{center}
  \includegraphics[scale=0.39,angle=0]{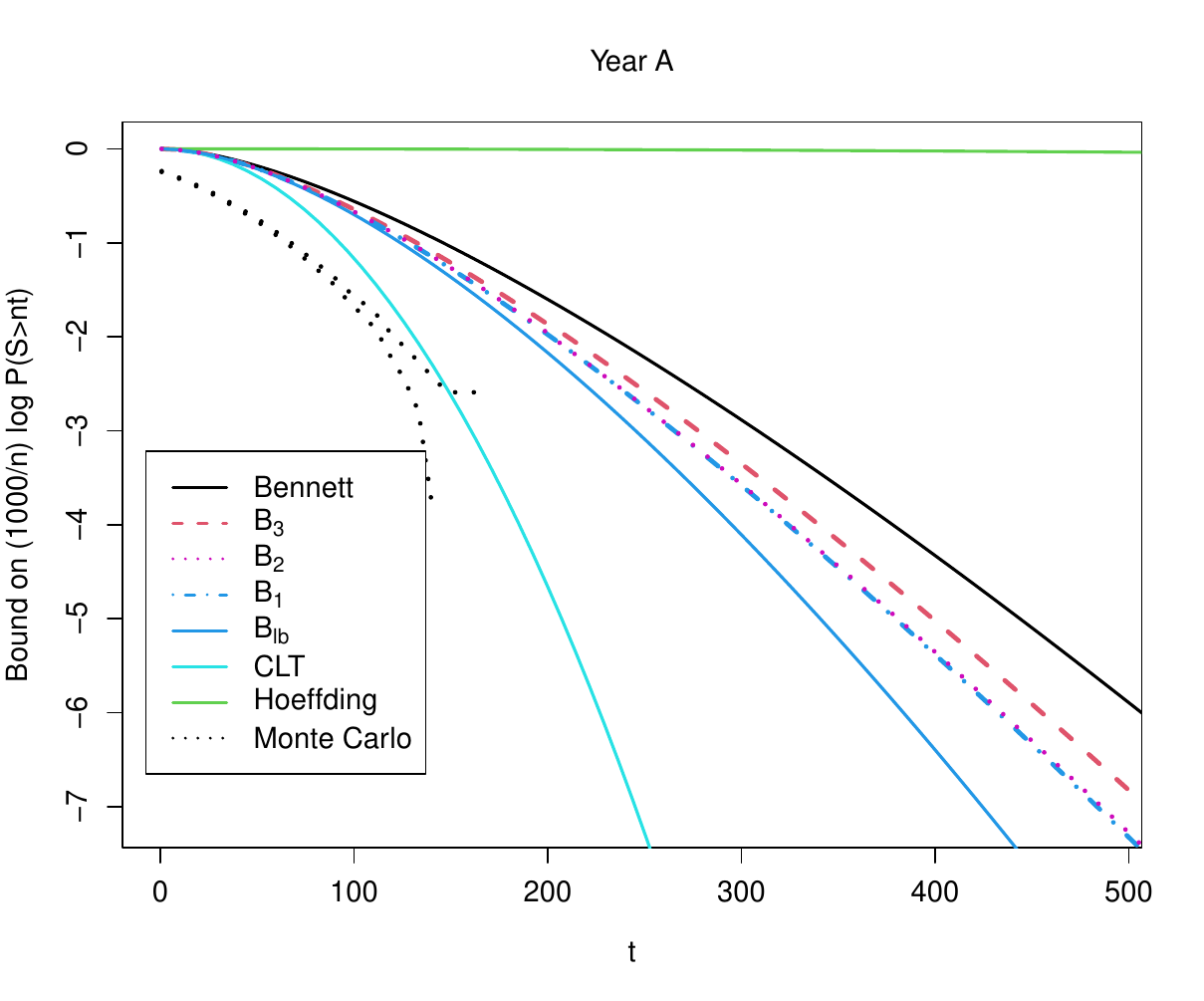}
  \includegraphics[scale=0.39,angle=0]{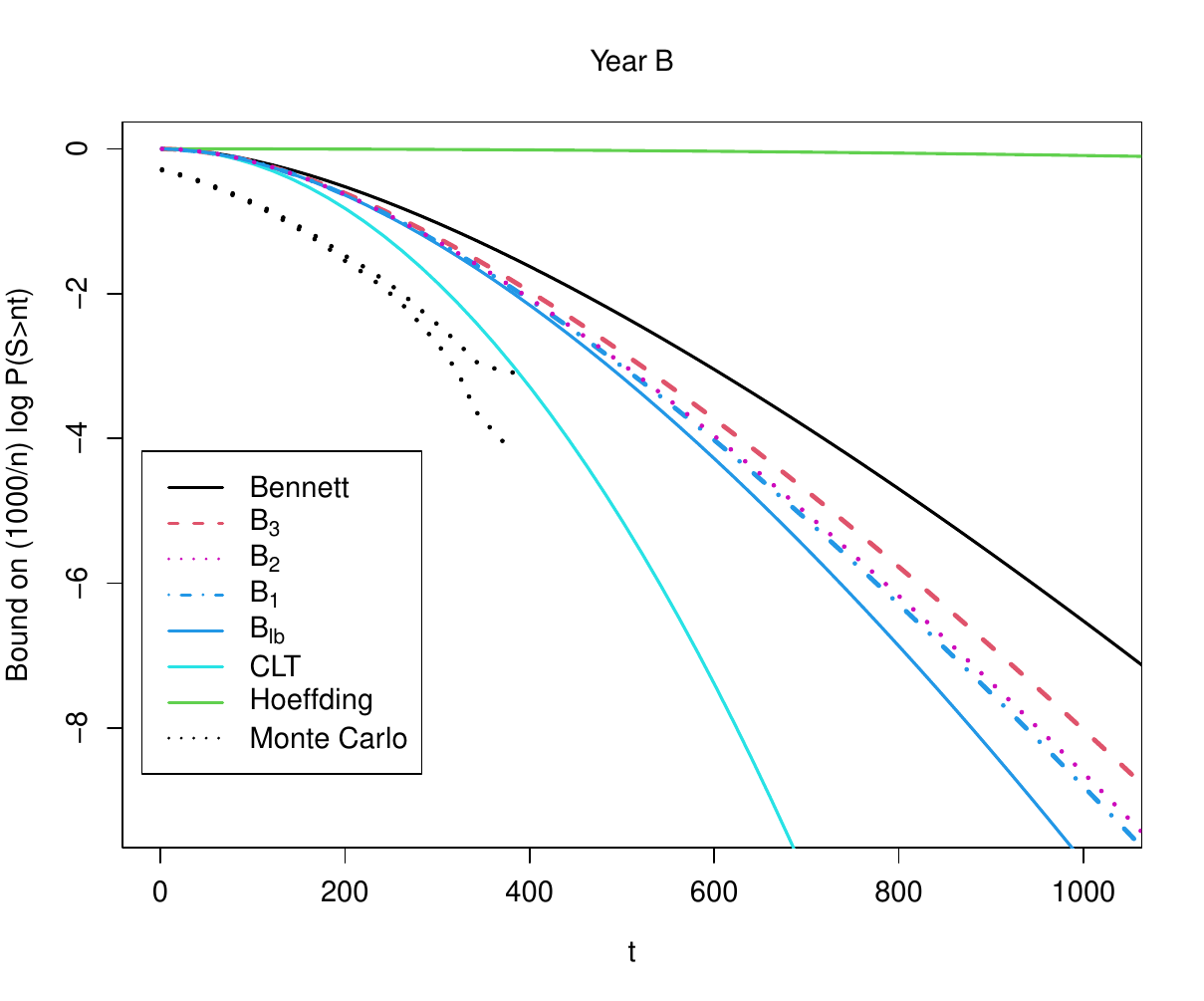}
  \includegraphics[scale=0.39,angle=0]{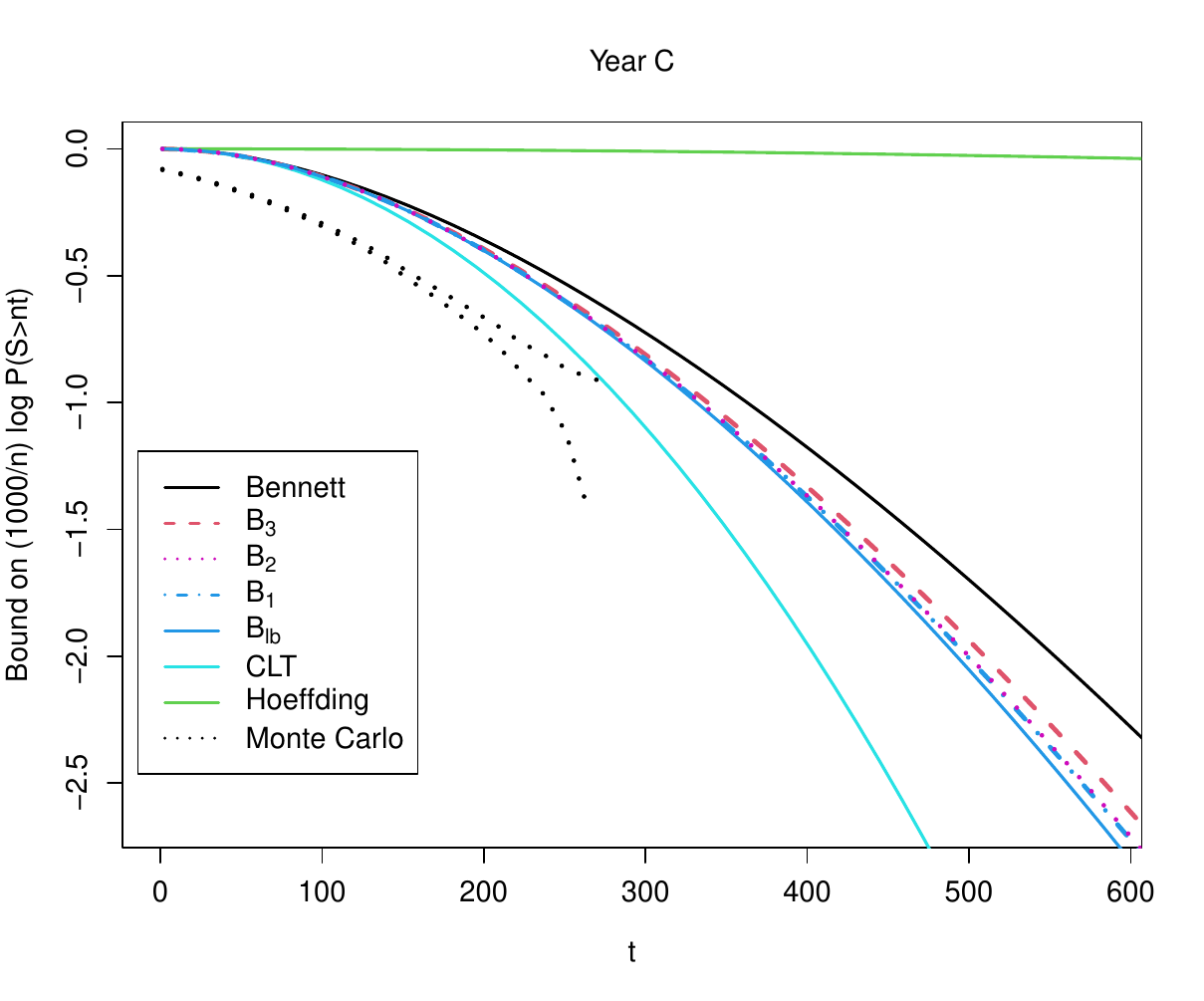}
\includegraphics[scale=0.39,angle=0]{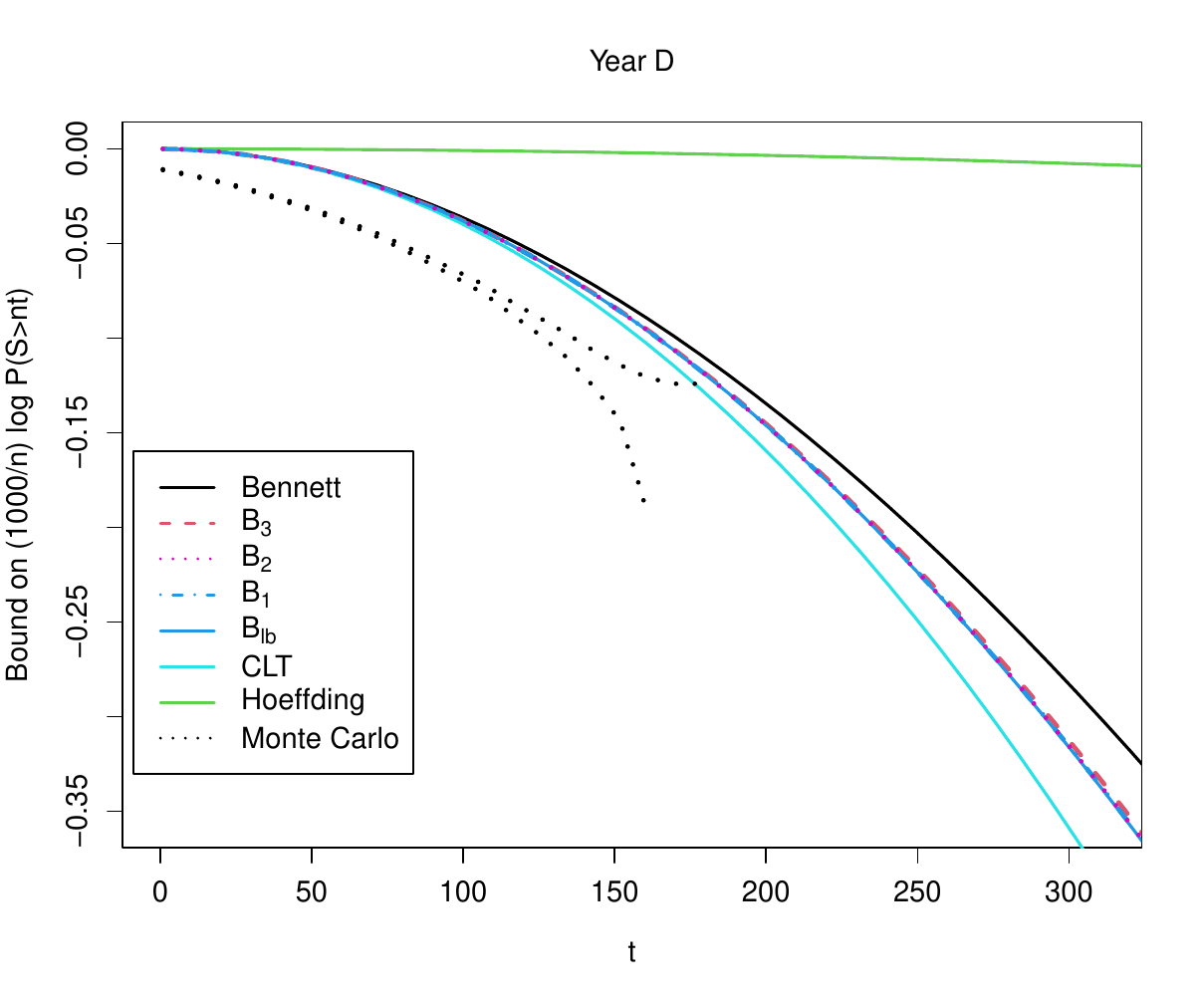} 
  \caption{Comparison of bounds on $(1000/n)\log \Prob{S_n\ge nt}$ plotted against $t$: Bennett (black solid); $B_1$ \eqref{eq.tightestP} in blue dot-dash; $B_2$ \eqref{eq.laxerP} in magenta dots and Equation \eqref{eq.laxestP}, denoted $B_3$, in red dashed. Other solid lines are: the large-$n$ approximation from the central limit theorem (cyan) and the Hoeffding bound (green). The dotted lines show 90\% CIs based on $20000$ Monte Carlo simulations from the true distribution. The four panels correspond to centred random variables, respectively for Years A, B, C and D. The $x$-axes of the panels are chosen so that the $y$-axes displays the same range of probabilities, from $1$ to approximately $10^{-10}$.
\label{fig.data.FourYears}
}
\end{center}
\end{figure}

For each of the years A, B, C and D, Figure \ref{fig.data.FourYears} considers the sums of the centred random variables and plots the Bennett and Hoeffding bounds and our bounds $B_1$ \eqref{eq.tightestP}, $B_2$ \eqref{eq.laxerP} and $B_3$ \eqref{eq.laxestP}, as well as the estimate from the central limit theorem and a 90\% confidence interval based on $20,000$ simulations from the true distribution. As expected, the Hoeffding inequality performs very poorly, and our bounds offer a substantial improvement on the Bennett bound. Bound $B_2$ is almost indistinguishable from $B_1$; for Years C and D, $B_1$ and $B_2$ are close to the best achievable by extensions of our method, and even for years A and B, the possible room for further improvement is half or less of that already achieved over the Bennett bound. Since we are especially interested in high return levels, it is years with expected losses between those of $C$ and $D$ that are most important. The approximation due to the central limit is poor, even though $n>200$ in all cases, because the individual loss distributions are highly skewed, with a high probability of no loss. With 20,000 Monte Carlo simulations it is only possible to measure probabilities down to about $1/5000$ with any accuracy, so the prediction-interval bounds cut off around this point.

An upper bound on $\frac{1}{n}\log \Prob{S\le nt}$ is obtained by applying our bounds to $-S$. Figure \ref{fig.data.FourYearsrev} in Appendix \ref{sec.addfigs} is the  equivalent of Figure \ref{fig.data.FourYears} but for the bounds on the lower tail probabilities. In all cases our bound is much closer to the Monte Carlo and the central limit-theorem estimates than for the upper tail, and is only slightly closer than the Bennett bound. As with the plots for the upper tail, the Bennett bound is much tighter than the Hoeffding bound. In contrast to $X_i$ which are typically strongly right skewed meaning that occasional large values are important, $-X_i$ is typically heavily skewed to the left, and so the upper tail of the sum has a behaviour much closer to that predicted by the central limit theorem.

\section{Fast, conservative bounds on return levels}
\label{sec.estimate.return.level}
We now describe the use of concentration inequalities to bound the $y$-year return level by bounding each sum $T_y$ in \eqref{eq.AddItAll} and how this leads to an approach that is several orders of magnitude  faster than the standard approach described in Section \ref{subsec.estimate.return.level}. 

Henceforth, we set $\mu_{y,e,r}=\alpha_{y,e,r}/(\alpha_{y,e,r}+\beta_{y,e,r})=\Expect{Z_{y,e,r}}$ and $X_{y,e,r}^{(s)}=b_{r}\{Z_{y,e,r}^{(s)}-p_{y,e,r}\mu_{y,e,r}\}/n_r^{(\mathrm{sub})}\le b_r\{1-p_{y,e,r}\mu_{y,e,r}\}/n_r^{(\mathrm{sub})}$, and consider the equivalent sum:
\begin{equation}
  \label{eqn.define.Sy}
  S_y=\sum_{e=1}^{n^{(\mathrm{ev})}_y}\sum_{r\in\mathcal{R}_e}\sum_{s=1}^{n_r^{(\mathrm{sub})}}X_{y,e,r}^{(s)},
  ~~~y\in\{1,\dots,n_y\}.
\end{equation}

\subsection{A new approach}
Some aspects of the new approach mimic the standard approach from Section \ref{subsec.estimate.return.level}: $M$ replicates for the total losses on each of the $n_y$ years are obtained. However, these replicates do not require individual simulations of every single loss at every single subrisk.

\begin{figure}
\begin{center}
  \includegraphics[scale=0.44,angle=0]{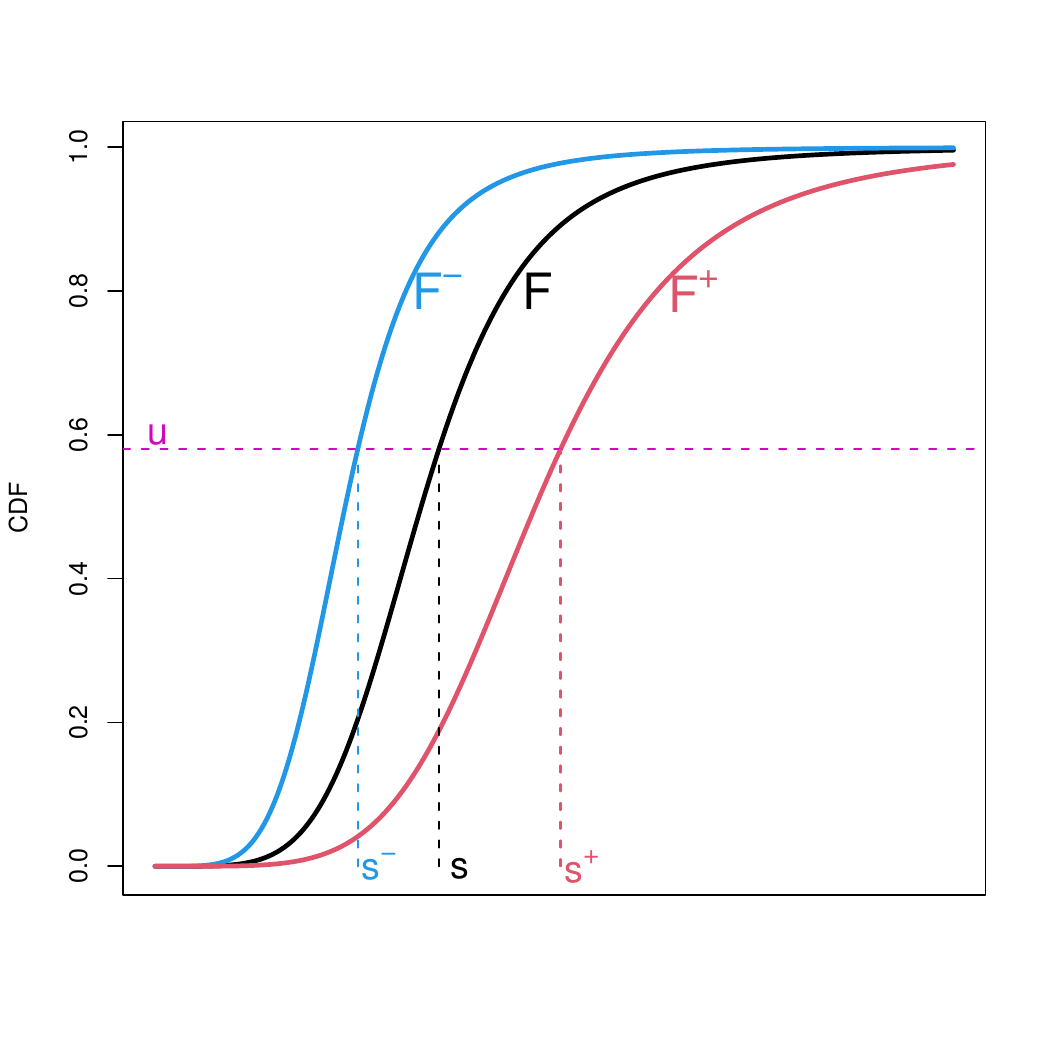}
  \caption{Illustration of a cumulative distribution function (cdf) $F$, the upper and lower bounds $F^-$ (blue) and $F^+$ (red) which are also cdfs, and the three resulting realisations, $s$, $s^-$ and $s^+$, for a particular value of $u$.
\label{fig.invCDF}
}
\end{center}
\end{figure}

Instead of a sample from the distribution of $S_y$ with (unknown and intractable) cumulative distribution function (cdf) $F_y$, we sample from the distribution of random variables $S_y^-$ and $S_y^+$ with known cdfs $F_y^-$ and $F_y^+$, which are ordered as follows:
\begin{equation*}
F_y^+(s)\le F_y(s)\le F_y^-(s).
\end{equation*}
Since both $F_y^-$ and $F_y^+$ are known, we may simulate from these simultaneously via the inverse cdf method: simulate a realisation $u$ from a $\mathsf{Unif}(0,1)$ random variable and then set $s_y^-$ and $s_y^+$ to be the solutions to $F_y^-(s_y^-)=u$ and $F_y^+(s_y^+)=u$. 
As shown in Figure \ref{fig.invCDF}, this implies that $s_y^-\le s_y\le s_y^+$, where $s_y=F^{-1}(u)$ is the hypothetical simulated value from the true distribution of $S_y$ that would be obtainable if its cdf were known. 

The distribution functions $F^-$ and $F^+$ arise directly from concentration inequalities as follows. Suppose that we have two concentration inequalities: $\Prob{S\le s}\le g_{lo}(s)$ and $\Prob{S \ge s}\le g_{hi}(s)$ where $g_{lo}(s)$ is increasing in $s$ and $g_{hi}(s)$ is decreasing in $s$. Defining $F^{-}(s):=g_{lo}(s)$ and $F^+(s):=1-g_{hi}(s)$, we have the required inequalities:
\begin{align*}
  F^+(s)&=1-g_{hi}(s)\le 1-\Prob{S\ge s}=F(s)=\Prob{S\le s}\le g_{lo}(s)=F^{-}(s).
\end{align*}

The following is repeated $M$ times: for each $y$, a realisation, $u_y$, from a $\mathsf{Unif}(0,1)$ distribution is sampled, and from this concurrent realisations from $s_y^-$ and $s_y^+$ are obtained via the inverse-cdf method. The corresponding realisations of the $k$-year return level are obtained from the order statistics: $q_k^-=s_{(n_y/k)}^-$ and $q_k^+=s_{(n_y/k)}^+$.

Since $s_y^-\le s_y\le s_y^+$ for each $y$, we have $q_k^-\le q_k\le q_k^+$, where $q_k$ is the return level that would have been obtained if we had sampled from the true distribution of $S_1, \dots,S_{n_y}$ via the inverse-cdf method.

The point estimate that would have been obtained can be bracketed between the mean (over the $M$ replicates) of the $q_k^-$ and the mean of the $q_k^+$. A conservative $95\%$ prediction interval can be obtained from the estimated $2.5\%$ quantile of the $q_k^-$ and the $97.5\%$ quantile of the $q_k^+$.

Typically $g_{lo}(s)=1$ for $s\ge \Expect{S_y}$ and $g_{hi}(s)=1$ for $s\le \Expect{S_y}$, so $s_y^-\le \Expect{S_y}$ and $s_y^+\ge \Expect{S_y}$, always. Firstly, however, the most important contributions to the estimate of a particular return level come from each $\Expect{S_y}$. Secondly, in terms of the effect of variability, when $s_y$ itself is simulated $M$ times, it is the exceptionally low or high values of $u_y$ and $s_y$ (for each $y$) that bracket its behaviour; the low values are well represented by the corresponding low values of $s_y^-$ and the high values by the high values of $s_y^+$; the low values of $s_y^+$ and the high values of $s_y^-$ are rarely important.

As written, each of our concentration inequalities provide an upper bound on $\Prob{S\ge s}$, which may be used as $g_{hi}(s)$. Applying the same type of bound to $-S$, with associated new values for each $c_i$, and hence $c_*$, leads to the bound $g_{lo}(s)\ge\Prob{-S\ge -s}=\Prob{S\le s}$.

By this method, the number of simulations is $\mathcal{O}(Mn_y)$. However, to obtain a single pair of samples, upper and lower bounds on the unobtainable `true' sample, we must solve $g_{lo}(s^-)=u=1-g_{hi}(s^+)$. These equations are not even tractable for Bennett's inequality \eqref{eq.Bennett}, and require a numerical solution. For $B_2$ \eqref{eq.laxerP} each step of the numerical solver involves an evaluation of Lambert's $W$, while for $B_1$ \eqref{eq.tightestP}, each step of the solver involves a numerical minimisation over $\lambda$.

\begin{table}
  
  \begin{center}
\caption{\label{table.CPUtime} CPU time (secs) for simulating from the small, standard and large portfolios. Timings are for the standard method and when using the concentration inequalities that give $B_1$ and $B_2$, and, finally, with $B_2$ using sampling-importance-resampling (SIR). Timings are reported for obtaining $M=100$ and $M=1000$ samples from the loss distribution for all $1000$ years. Each complete simulation was repeated 10 times; the table reports the mean and standard deviation of the timings. }

%\begin{tabular}{l|cccccc}
%  &\multicolumn{2}{c}{\textbf{Small Portfolio}}&\multicolumn{2}{c}{\textbf{Standard Portfolio}}&\multicolumn{2}{c}{\textbf{Large Portfolio}}\\
%  &$M=100$&$M=1000$&$M=100$&$M=1000$&$M=100$&$M=1000$\\
%  \hline
%  \textbf{Std. Method}\\
%  Set up&\multicolumn{2}{c}{$0.75$ ($0.15$)}&\multicolumn{2}{c}{$4.22$ ($0.31$)}&\multicolumn{2}{c}{$23.4$ ($0.63$)}\\
%  Simulations&$49.2$ ($0.97$) &$511$ ($25.3$)&$298$ ($6.3$)&$2870$ ($96.4$)&2719 (100)&25796 (TODO)\\
%  \hline
%  \textbf{Concentration}\\
%  Set up&\multicolumn{2}{c}{$1.44$ ($0.052$)}&\multicolumn{2}{c}{$6.03$ ($0.132$)}&\multicolumn{2}{c}{$47.2$ ($1.52$)}\\
%  Sim. direct $B_1$&196 (6.30)&t (sd)& 194 (13.2)&1740(253)&200 (25.7)&2006 (399)\\
%  Sim. direct $B_2$&98.2 (3.82)&965 (142)&108 (9.18)& 1070 (152)&116 (17.2)&1119 (214)\\
%  Sim. SIR $B_2$&1.35 (0.056)&2.82 (0.131)&1.47 (0.239)&2.79 (0.093)&1.40 (0.035)&2.90 (0.068)
  % for M=100, 8 around 1.35 and 2 around 1.9.
%\end{tabular}

\begin{tabular}{c|r|r|rr|r}
  Portfolio&M&Std. Method&$B_1$ (direct)&$B_2$ (direct)&$B_2$ (SIR)\\
  \hline
  small&100&49.2 (0.97)&196 (6.30)&98.2 (3.82)&1.35 (0.056)\\
  standard&100&298 (6.3)&194 (13.2)&108 (9.18)&1.47 (0.239)\\
  large&100&2719 (100)&200 (25.7)&116 (17.2)&1.40 (0.035)\\
  \hline
  small&1000&511 (25.3)&1845 (235)&965 (142)&2.82 (0.131)\\
  standard&1000&2870 (96.4)&1740 (253)&1070 (152)&2.79 (0.093)\\
  large&1000&26221 (352)&2006 (399)&1119 (214)&2.90 (0.068)
  \end{tabular}

  \end{center}

  \end{table}

Table \ref{table.CPUtime} compares the timings for simulation via the standard method and using the inverse cdf method with $B_1$ and $B_2$ for $1000$ years of floods using our standard portfolio, which has $10^5$ risks and approximately $1.8\times 10^6$ subrisks, both using $M=100$ and $M=1000$; this is repeated for the small portfolio from which our standard portfolio was derived (see Section \ref{sec.stdPortfolio}) and for the large portfolio, ten times the size of the standard portfolio. Before simulations can proceed, each method has a set up cost, where the event and portfolio information is read in. For the concentration-based methods, summaries for each year, such as $n$, $\sigmasqbar$ and $K$, are also calculated. These pre-simulation \emph{set up} times for the standard method were, in seconds, 0.75 (0.97) for the Small Portfolio, 4.22 (0.31) for the Standard Portfolio and 23.4 (0.63) for the Large Portfolio. The corresponding set-up times for the three portfolios for the concentration methods were: 1.44 (0.052), 6.03 (0.132) and 47.2 (1.52).

Recall that our simulations are for $1000$ years of flood events. To give more reliability on estimates of the high return levels it is more typical to simulate from $10000$ years of events. The full UK JBA Flood Event Set has an average of $10.6$ events per year \cite[]{JBAthree}. Also, our standard portfolio has about $10^5$ risks, whereas a large portfolio, for the UK or for the whole of Western Europe, for example, might have $10^6$ or more risks. This means that simulations for a large portfolio can take 24 hours or more to complete in practice.

As expected, the CPU time for the standard method scales in proportion to $M$ and, except for the small portfolio, where the relative cost of other overheads is larger, in proportion to the portfolio size. Again, as expected, for all concentration-based methods, there is little dependence on the portfolio size, but  direct simulation using $B_1$ or $B_2$ has a cost proportional to $M$. For $M=1000$, simulating using $B_1$ takes about $60\%$ of the CPU time that the standard method takes, and simulating $B_2$ takes about $37\%$ of the CPU time; typically, there is less of an improvement when $M=100$, and more improvement the larger the portfolio size. Set up costs for $B_1$ and $B_2$ are typically twice those for the standard method, but these are dwarfed by the simulation costs.

 The table also includes times for simulating from $B_2$ via an importance sampling-resampling method. For the standard portfolio, this method is a factor of around $200$ times faster than the standard method when $M=100$, with a factor of around $1000$ when $M=1000$. These ratios improve by a further factor of $10$ for the larger portfolio. Interestingly the cost of the SIR simulations only doubles when $M$ increases by a factor of $10$, presumably due to the vectorisation of the simulation, transformation and weighting operations. We now describe this importance-sampling method.

\subsection{Importance sampling-resampling}

Importance sampling obtains an approximate, weighted sample, $\{(x_i,w_i)\}_{i=1}^M$, from a distribution with a density $f$ by simulating a sample, $x_1,\dots,x_M$, from a distribution with a density $q$, called the proposal density, setting  $w^*_i=f(x_i)/q(x_i)$ for $i=1,\dots,n$ and finally setting $w_i=w^*_i/\sum_{j=1}^n w^*_j$. The Monte Carlo representation of the intended distribution is then
$
\sum_{i=1}^M w_i \delta(x-x_i),
$
where $\delta$ is Dirac's $\delta$-function. Sampling $M$ times from this weighted approximation to $f$ gives an unweighted sample which can be used to approximate $f$.

Throughout this exposition we consider simulation from $F^+$; simulation from $F^-$ proceeds analogously. Our proposal distribution is based on Bernstein's inequality:
\[
\frac{1}{n}\log    \Prob{S_n\ge nt}
\le -\frac{\frac{1}{2}t^2}{\sigmasqbar+\frac{1}{3}c_* t}.
\]
From this,
\[
P(S_n \ge t')\le \exp\left[-\frac{\frac{1}{2}{t'}^2}{n\sigmasqbar+\frac{1}{3}c_*t'}\right]=:S_{Ber}(t').
\]
Bernstein's inequality is weaker than Bennett's inequality and can be derived directly from it \cite[e.g.,][Section 2.7]{BoucheronLugosiMassart2013}. Unlike Bennett's inequality, however, it is straightforward to sample from the distribution implied by Bernstein's inequality, solving $S_{Ber}(t')=u\sim \mathsf{Unif}(0,1)$ to obtain
\[
t'=\left[\left\{\frac{1}{3}c_*\log u\right\}^2-2n\sigmasqbar \log u\right]^{1/2}-\frac{1}{3}c_*\log u.
\]
Differentiation of $S_{Ber}(t')$ gives the proposal density:
\[
q(t')=\frac{(n\sigmasqbar+\frac{1}{6}c_*t')t'}{(n\sigmasqbar+\frac{1}{3}c_*t')^2}S_{Ber}(t').
\]
%\[
%\frac{1}{n}\log    \Prob{S_n\ge nt}
%\le -\frac{\sigmasqbar}{c_*^2} \times \frac{\{c_* t/\sigmasqbar\}^2}{2+ 2\{c_* t/\sigmasqbar\}/3}.
%\]
%From this $P(S_n \ge t')\le \exp\left[-n \frac{\sigmasqbar}{c_*^2}\frac{v^2}{2+2v/3}\right]=:S_{Ber}(t')$ with
%\[
%v=\frac{c_* t'}{n\sigmasqbar}.
%\]
%Bernstein's inequality is weaker than Bennett's inequality and can be derived directly from it \cite[e.g.,][Section 2.7]{BoucheronLugosiMassart2013}. Unlike Bennett's inequality, however, it is straightforward to sample from the Bernstein approximation, solving $S_{Ber}(t')=u$ to obtain
%\[
%t'=\left[\left\{\frac{1}{3}c_*\log u\right\}^2-2n\sigmasqbar \log u\right]^{1/2}-\frac{1}{3}c_*\log u.
%\]
%Differentiation of $S_{Ber}(t')$ gives the proposal density:
%\[
%q(t')=\frac{2}{c_*}\frac{v(2+v/3)}{(2+2v/3)^2}S_{Ber}(t').
%\]
We consider $S_{BS}:=\exp\{nB-\lambda t'\}$ with $B$ as defined in \eqref{eq.new.mgf.bound.tightest}, for which
\[
f_{BS}(t')=\left[\left\{n\frac{\md B}{\md \lambda}-t'\right\} \frac{\md\lambda}{\md t'} -\lambda\right]S_{BS}(t').
\]
It is straightforward to show that $({\md}/{\md \lambda}) \{\lambda^j f_j(\lambda c_*)\}=\lambda^{j-1}f_{j-1}(\lambda c_*)$, from which,
\[
\frac{\md B}{\md \lambda}= \lambda(\sigmasqbar-K)+K\lambda f_1(\lambda c_*)-K_1c_*^2\lambda^3f_3(\lambda c_*).
\]
We have not been able to obtain $\md \lambda /\md t'$ corresponding to the numerical solution for $\lambda$ in \eqref{eq.tightestP}; however, for \eqref{eq.laxerP}, we show in Appendix \ref{eqn.sec.derive.dlambydt} that
\[
\frac{\md\lambda}{\md t'}=\frac{1}{n(\sigmasqbar-K)}\times \frac{1}{1+W},
\]
where
\begin{equation}
  \label{eq.W.for.dlambydt}
W=\calW\left(\frac{K}{\sigmasqbar-K}\exp\left[\frac{K+t'c_*/n}{\sigmasqbar-K}\right]\right).
\end{equation}
The importance-sampling-resampling method, requires only a single evaluation of $\calW$ per sample and leads to the impressive speed up illustrated in Table \ref{table.CPUtime}. Figures \ref{fig.numerical.expts} and \ref{fig.data.FourYears}  show that the bound $B_2$ from \eqref{eq.laxerP} is close to $B_1$ from \eqref{eq.tightestP}. We, therefore, recommend using importance-sampling-resampling with bound $B_2$. 

Strictly speaking, the unweighted sample obtained by sampling-importance-resampling is not a sample from $f$ but its distribution tends to that of a sample from $f$ as $M\uparrow \infty$. To improve the accuracy of the sample we use residual resampling \cite[e.g.,][]{DoucetJohansen2008}.
Figures \ref{fig.logProb.directvSIR}, \ref{fig.rtnlvl.directvSIRa} and \ref{fig.rtnlvl.directvSIRb} in Appendix \ref{sec.cmpISwithDirect} demonstrate empirically that in our examples the samples are visually indistinguishable from a sample from $f$.

\subsection{Numerical comparison}

We now compare results from the standard method with those from our importance sampling-resampling method using bound $B_2$. Estimating a 95\% prediction interval using only $M=100$ realisations leads to considerable Monte Carlo variability, so to mitigate for this source of variability, we use $M=1000$ throughout this comparison.

\begin{figure}
\begin{center}
  \includegraphics[scale=0.39,angle=0]{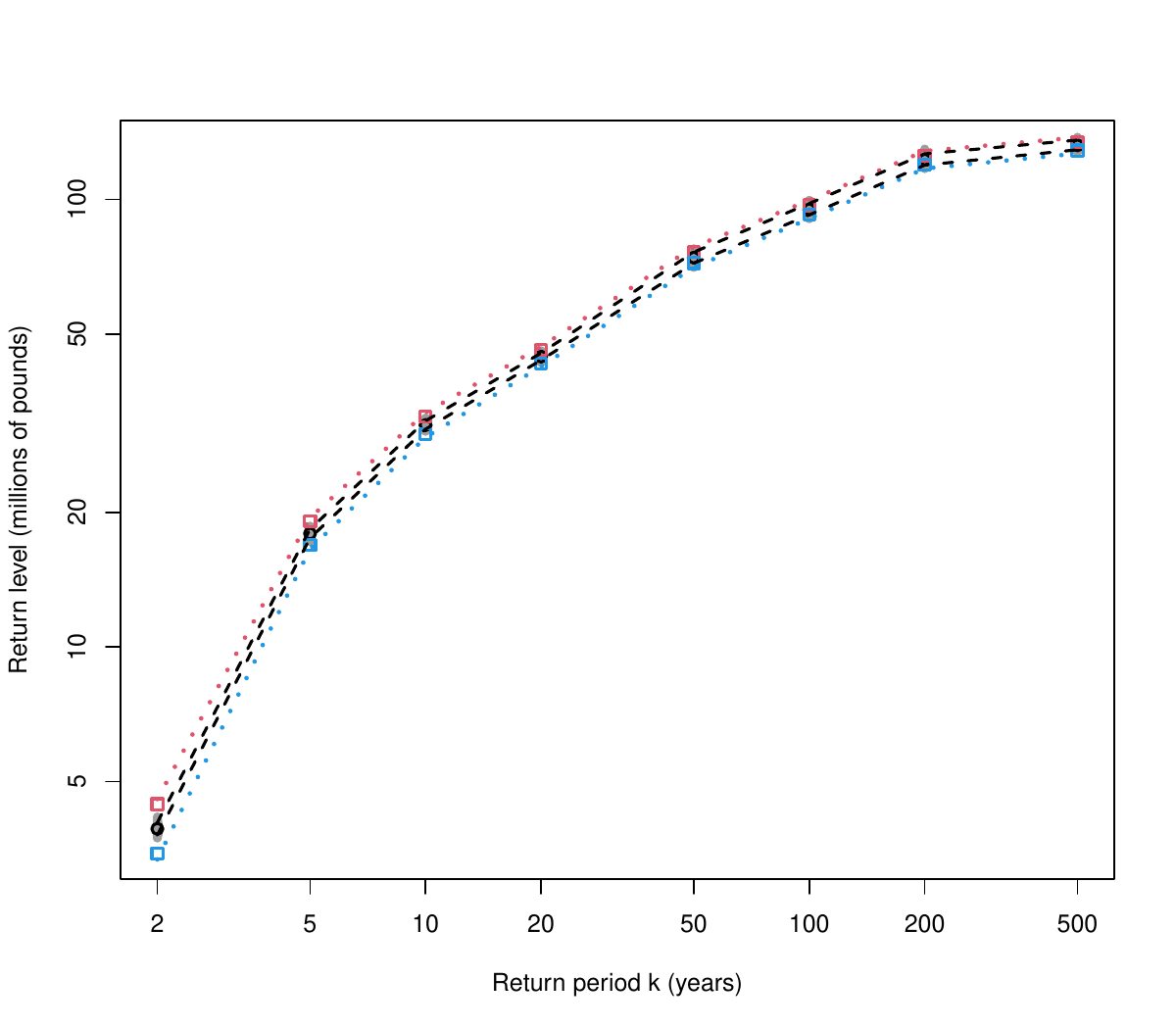}
  \includegraphics[scale=0.39,angle=0]{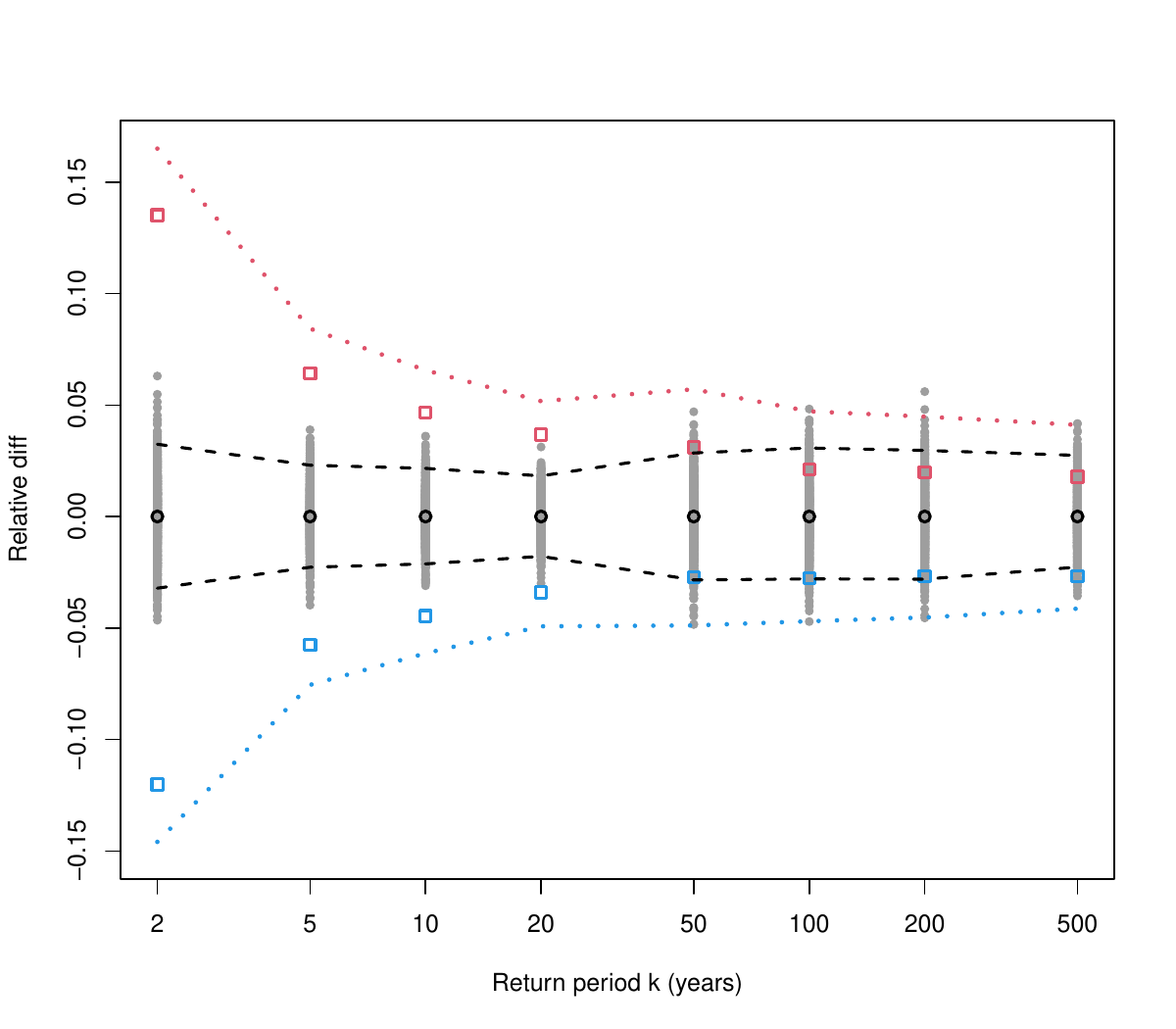}
  \caption{Return level point estimates and 95\% predictive bounds via the standard method (in black) and via sampling importance resampling using bound $B_2$ for $F^+$ and $F^-$. The point estimate and upper end of the prediction interval when using $F^+$ appear in red; when using $B_2$ for $F^-$, the corresponding point estimate and lower predictive bound appear in blue. The first plot shows the raw return level predictions. For each return level, the second plot shows all values, $v$ from the first plot relative to the point estimate, $s$, for that return level obtained from the standard method: \emph{i.e.}, $(v-s)/s$. On the second plot, the $1000$ realisations from the standard method appear in grey. Logarithmic spacing is used for both axes in the top plot and for the x-axis in the bottom plot.
\label{fig.rtnlvlcomp}
}
\end{center}
\end{figure}

Figure \ref{fig.rtnlvlcomp} shows the estimated $k$-year return levels for $k=2$, $5$, $10$, $20$, $50$, $100$, $200$, $500$. The first plot shows that the estimates obtained from $B_2$ closely track the estimates from the standard method. The second plot shows that the intervals when using $B_2$ are wider than those obtained by the standard method, but that they get narrower as the return level increases. From the 10-year return level upwards, both the upper and lower point estimates are within 5\% of the point estimate from the standard method, reducing to 2\% for the 100-year and higher return levels. Similarly, for all return levels from the 10-year level upwards, the estimates of the prediction bounds are within around 5\% of the bounds obtained from the standard method, reducing to a few percent from the 100-year level upwards. Errors in estimating the two-year return level, especially, are more substantial. 

Table \ref{table.relCIwidth} shows for each return level the relative width of the 95\% prediction interval obtained via our conservative methodology to that of the interval obtained using the standard method. This ratio is large for the two-year return level and decreases with increasing return level. For the 5-year return level it is less than a factor of $3.5$ and for the 50-year return level and above it is less than a factor of $2$. From an insurance perspective the total claim is of most importance, the additional ``noise'' from our method is tiny compared with the expected loss. For example with the standard method it is estimated that an annual loss of approximately $123\pm 3.5$ million pounds will be exceeded once in every 200 years, whereas with our approach we obtain a 95\% prediction interval of  $123\pm 5.5$ million pounds.

\begin{table}
  \begin{center}
    \caption{\label{table.relCIwidth}Relative width of the 95\% prediction interval for each return level from the concentration approach compared with the standard method. Bootstrapped standard errors appear in brackets.}
\begin{tabular}{c|cccc}
  Return level&2&5&10&20\\
  Relative width&5.099 (0.123)&3.448 (0.102)&2.928(0.091)&2.713 (0.066) \\
  \hline
  Return level&50&100&200&500\\
  Relative width&1.880 (0.061)&1.624 (0.054)&1.577 (0.063)&1.546 (0.062)
\end{tabular}
  \end{center}
\end{table}

Figure \ref{fig.rtnlvlcompB} in Appendix \ref{app.more.rtn.lvls} repeats the second plot in Figure \ref{fig.rtnlvlcomp} but with a different random seed for the standard portfolio, and then for the large portfolio. Both show the same overall pattern that is seen in Figure \ref{fig.rtnlvlcompB}, but we highlight two changes with the larger portfolio. Firstly, the relative differences are smaller, as would be expected since yearly expected losses are proportional to $n$, the portfolio size, whereas standard deviations are proportional to $\sqrt{n}$. Furthermore, the discrepancies between the predicted CIs using $B_2$ and those using the standard method are slightly smaller. 

Figure \ref{fig.data.FourYears} shows a substantial deviation between the loss from $B_2$ and $B_{lb}$ for the year (A) with lower quartile expected loss. The upper tails of years with the lower expected losses will impact the lower return levels and we conjecture that for these years, a tighter concentration bound, as discussed in Section \ref{sec.discuss.conc} and Appendix \ref{app.tighter.still}, might lead to narrower intervals for the lower return levels.

\subsection{Sensitivity study}

The parameters $\alpha_{y,e,r}$ and $\beta_{y,e,r}$ are obtained from the spatial flood event simulated from the multivariate extremes model as follows. The extremes model provides maximum river flow levels at each measurement site; the hydraulic model then converts these to water depths over a fine grid of points in the region; finally, the water depths in the vicinity of a particular risk are then converted to a probability of flooding for that risk and an expectation and variance of the damage ratio (the fraction of the insured value that will be claimed given that there will be a claim) via \emph{depth-damage curves} --- curves specifying the damage ratio as a function of depth for different categories of risk. The parameters $\alpha_{y,e,r}$ and $\beta_{y,e,r}$ are obtained by matching the expectation and variance of the damage ratio distribution.

As stated in \cite{MFGCG2020}, \cite{Freni2010} and \cite{JKABCB20212}, the use of depth-damage curves involves great uncertainty. The actual damage to a property also depends on factors such as the duration of a flood, the water velocity, suspended debris, sewer and drainage blockages and the amount of warning time.

As the compute time of our simulation methodology is orders of magnitude smaller that of than the standard methodology, this enables sensitivity studies which were previously infeasible. We illustrate this by imagining perturbations to the expected damage ratio for each risk. Specifically if the expected damage ratio for a particular $(y,e,r)$ combination is perturbed by $\delta$, we then alter $\alpha_{y,e,r}$ and $\beta_{y,e,r}$ to $\alpha'_{y,e,r}$ and $\beta'_{y,e,r}$ so that the expectation $\mu_{y,e,r}$ changes as follows:
\[
\mu'_{y,e,r}=\frac{\alpha'_{y,e,r}}{\alpha'_{y,e,r}+\beta'_{y,e,r}}
=
(1+\delta)\frac{\alpha_{y,e,r}}{\alpha_{y,e,r}+\beta_{y,e,r}}
=
(1+\delta)\mu_{y,e,r}.
\]
For all risks across all events, $\mu_{y,e,r}<0.95$; however, if ever $(1+\delta)\mu_{y,e,r}>0.95$, we set $\mu'_{y,e,r}=0.95$. In either case we keep $\alpha_{y,e,r}+\beta_{y,e,r}$ fixed as this has an interpretation as the total amount of information.

\begin{table}
  \begin{center}
\caption{\label{table.sensitivity} Statistics for the 200-year return level. Estimated 2.5\% and 50\% quantiles from the conservative lower bounds and the 50\% and 97.5\% quantiles from the conservative upper bounds in million £s to 4s.f. for different portfolio scenarios. For scenarios P3 and P4 the estimates are from a single pooled sample of all of the $R\times M$ simulations of the return level.}

\begin{tabular}{c|c|c|c|c}
  Scenario & Lower 2.5\% quantile & Lower median & Upper median & Upper 97.5\% quantile\\
  \hline
  P0 & 117.3 & 119.7 & 125.1 & 128.6 \\
  P1 & 123.3 & 125.7 & 131.4 & 134.8 \\
  P2 & 111.3 & 113.7 & 119.0 & 122.2 \\
  P3 & 117.3 & 119.8 & 125.2 & 128.6 \\
  P4 & 116.5 & 120.1 & 125.7 & 129.6
  \end{tabular}

  \end{center}
  \end{table}

Naming our standard portfolio P0, we consider four perturbation scenarios for this portfolio:
\begin{itemize}
\item[P1] For all risks the expected damage ratio is perturbed by a factor of $+0.05$.
\item[P2] For all risks the expected damage ratio is perturbed by a factor of $-0.05$.
\item[P3] For each risk an unbiased coin is flipped and the expected damage ratio is perturbed by $\pm 0.05$ depending on the outcome.
\item[P4] For each risk an unbiased coin is flipped and the expected damage ratio is perturbed by $\pm 0.25$ depending on the outcome.  
\end{itemize}

In the case of P3 and P4, if $(1+\delta)\mu_{y,e,r}>0.95$ we set $\mu'_{y,e,r}=0.95$ with a probability of $0.5$ and $\mu'_{y,e,r}=2\mu_{y,e,r}-0.95$ otherwise.

For each of the deterministic scenarios P1 and P2 we simulated $M=1000$ sets of losses for each year, leading to samples of $M$ conservative upper values and $M$ conservative lower values for each return level. The first three columns in Table~\ref{table.sensitivity} show the estimated 2.5\% and 50\% quantiles from the conservative lower bounds and the 50\% and 97.5\% quantiles from the conservative upper bounds for P0, P1 and P2. Unsurprisingly, the quantiles for P1 are around a factor of 0.05 larger than those for P0, while those for P2 are a factor of 0.05 smaller.

For the random scenarios P3 and P4, we simulated $R=100$ different perturbations of the standard portfolio and for each of these we simulated $M=1000$ sets of conservative upper bounds on the losses for each year, leading to $R$ sets of $M$ values for each return level. We repeated this procedure for the same $R$ perturbations, but simulating, instead, conservative lower bounds. For a given return level, these simulations can be viewed in two useful ways.

\begin{figure}
\begin{center}
\includegraphics[scale=0.6]{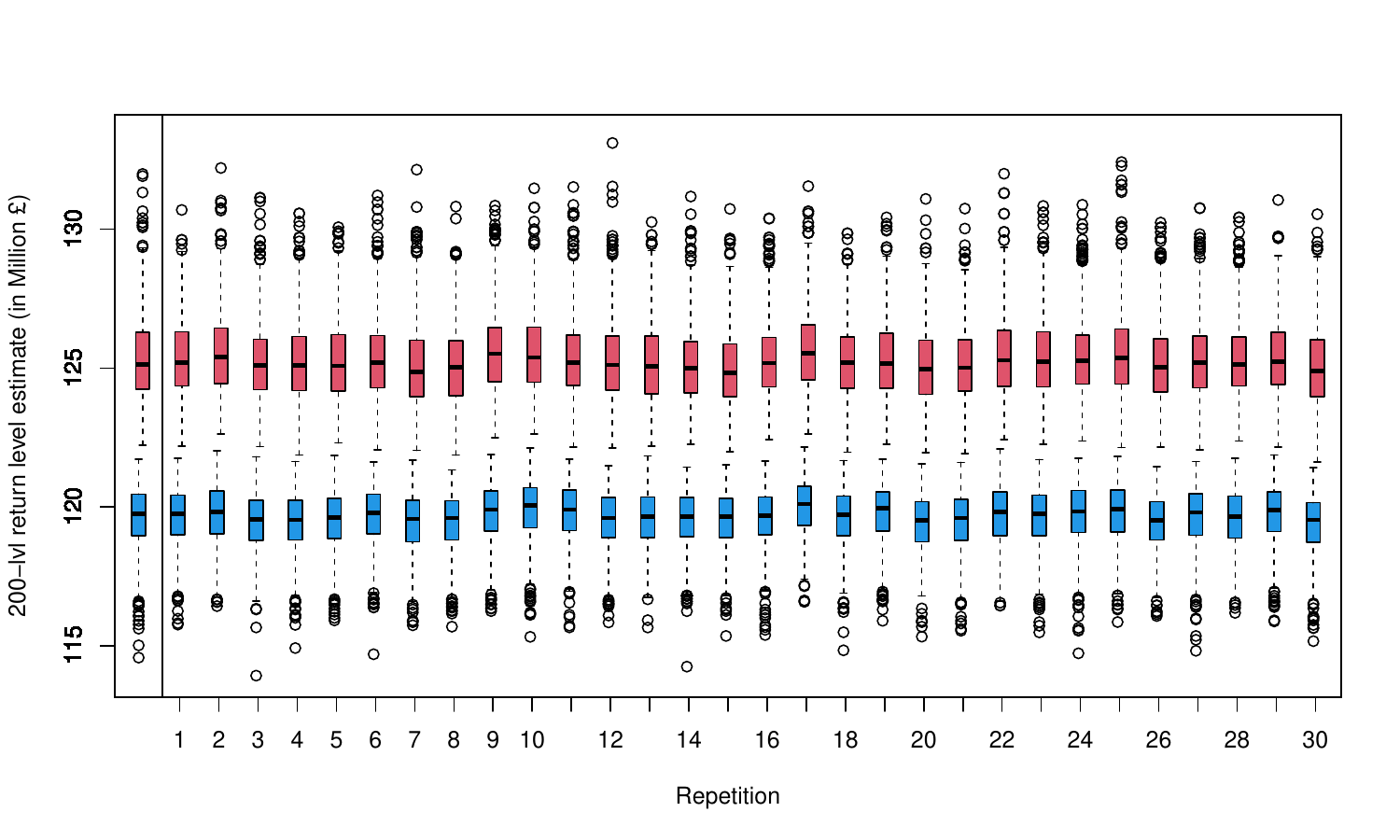}
\includegraphics[scale=0.6]{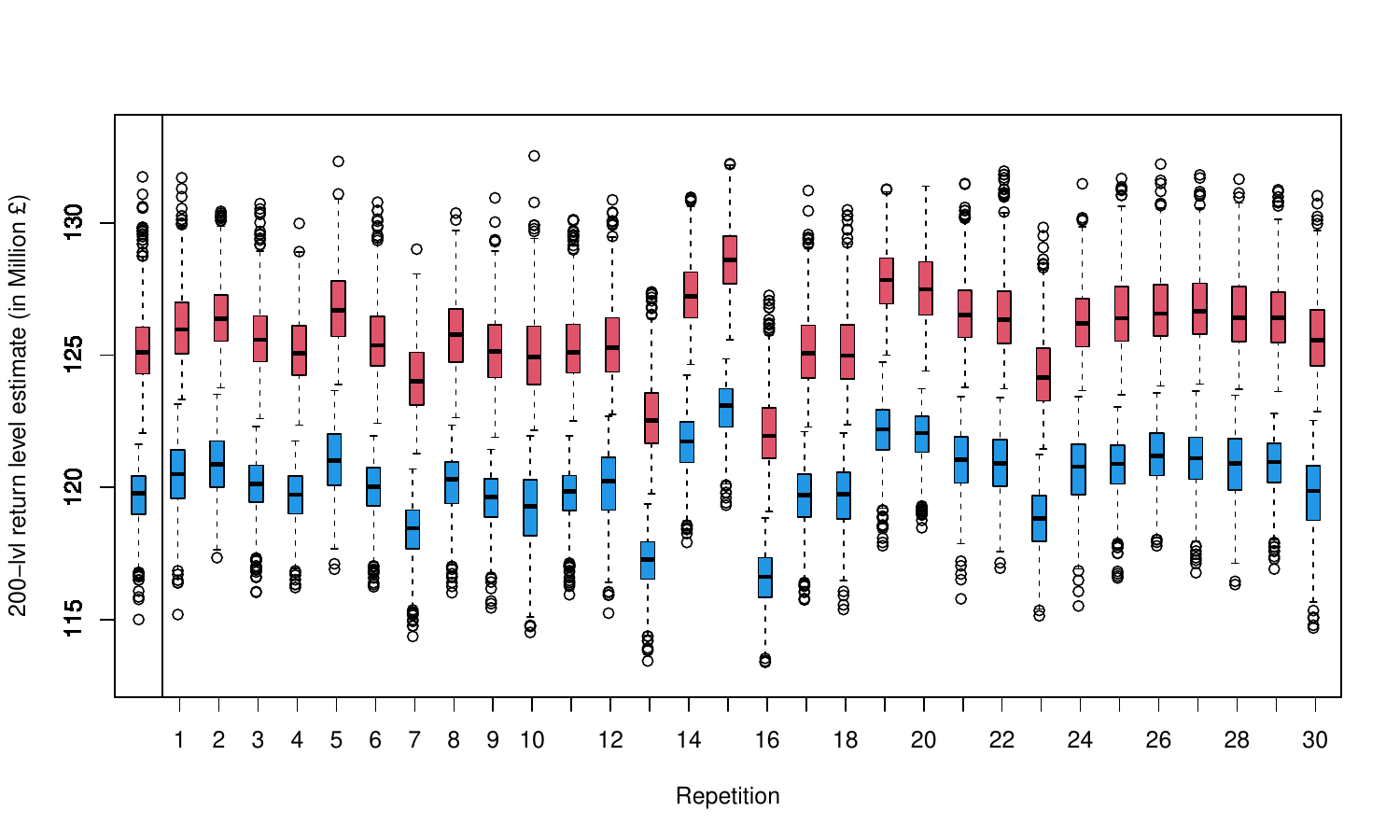}
  \caption{Boxplots of the 1000 simulated 200-year return levels for the portfolio scenarios P3 (top panel) and P4 (lower panel). The red/blue boxplots show the estimates based on the upper/lower conservative bound. The first 30 boxplots of 100 replicates are shown. The two leftmost are boxplots of return level estimates from our original portfolio P0 for comparison.
\label{fig.boxplots}
}
\end{center}
\end{figure}

Firstly, each of the $R$ replicates shows an alternative possibility that might have been obtained if the expected damage ratios had been calculated slightly differently, perhaps taking flood duration and the amount of warning into account. Figure~\ref{fig.boxplots} examines the simulated 200-year return levels and shows the first $30$ of the R for each of P3 (top panel) and P4 (lower panel). With P3 there is very little variability from one possible portfolio to the next, suggesting that perturbations of the order of 5\% make little difference to the estimated quantiles; in P4, where the perturbations are much more substantial, the discrepancies are more visible but are still much smaller than the variability in the $M$ simulations within any scenario.

Secondly, we can view the perturbations of the portfolios as representing extra sources of uncertainty, such as the duration of the flood, the amount of warning, or the water velocity, which we would like to account for in our estimation of return levels. Thus we may pool all of the $R\times M$ simulations of a particular return level into a single large sample of that return level; we do this separately for the upper conservative simulations and for the lower conservative simulations. Quantiles from these samples then represent our estimates taking this additional uncertainty into account and comprise the final two columns of Table~\ref{table.sensitivity}. We see that the additional uncertainty in the damage ratio makes little difference to the estimated quantiles, and can understand this through the relatively small between-simulation variability, $\sigma^2_b$, compared to the within-simulation variability, $\sigma^2_w$, visible in the box plots of Figure~\ref{fig.boxplots}. To quantify this, for P3, $\sigma^2_b/\sigma^2_w$ is approximately 1/250 and 1/400, respectively for the upper and lower bounds, whereas for P4 the ratios are approximately 1/8 and 1/11. Further experiments show that these fractions are generally smaller for lower return levels. To achieve a ratio of $1$ we found it necessary to examine the 500-year return level and increase delta to the unrealistically high value of $0.7$. Appendix~\ref{app:sensitivity} provides analogous summary statistics, boxplots and ratios for the 20-year return level. 

The overall message with regard to our portfolio is that it is vital that estimates of the expected damage ratio are as close as possible to being unbiased, but it is less important to account correctly for all the uncertainty. The story may be different with other portfolios: our methodology will facilitate analogous simulation studies for these.

\section{Discussion}
\label{sec.discussion}
Repeated simulations of individual losses for each subrisk over an entire portfolio for each of $1000$ or $10000$ years worth of river flood events is computationally very costly. We have described a new methodology for simulating bounds on yearly losses which is based upon the cumulative distribution functions implied by a new concentration inequality. The upper and lower bounds on the simulated yearly losses reduce the computational cost of estimating return levels and their uncertainty by orders of magnitude. %We have demonstrated the utility of this in a short sensitivity study.
We have demonstrated the utility of this in a short sensitivity study which found that for our portfolio, any bias in the expected damage ratios carries through directly to the estimated return levels and their quantiles; however even allowing for quite substantial additional (unbiased) uncertainty in this expectation makes little difference to the estimates.

The same methodology could be applied anywhere that many replicate simulations of sums of large numbers of independent random variables are required. For example, for other insurance events, such as coastal flooding or surface-water flooding, where the final stage of simulation is conditional on the output of an extreme-value model. 

Although the flood-event extreme-value model itself is not the subject of this article, it is an important part of the overall catastrophe-modelling (CAT) framework. There are opportunities for improving this, too, and these are discussed briefly in Appendix \ref{sec.EVTmodel}, where the general CAT framework is also overviewed.

We first discuss the new concentration inequalities, then the new simulation methodology. Also, at first glance, two natural potential alternatives to our procedure suggest themselves. Appendix \ref{sec.alternativeApproaches} describes both possibilities and explains why neither is suitable for our purposes.

\subsection{The concentration inequalities}
\label{sec.discuss.conc}
Our new inequalities, \eqref{eq.tightestP} to \eqref{eq.laxestP}, are extensions of Bennett's inequality \eqref{eq.Bennett}. As with Bennett's inequality, the support of individual variables to be summed has a finite upper bound, and the individual variances are taken into account. In contrast to Bennett's inequality which employs a single upper bound on the supports of all the variables, our bounds account for the individual supports. Our tightest bound \eqref{eq.tightestP} requires numerical optimisation, but both \eqref{eq.laxerP} and \eqref{eq.laxestP} are tractable via Lambert's W-function.

Alternative tightenings of Bennett's inequality are possible in cases where $\sigma_i/c_i$ has a non-trivial lower or upper bound. For the former, see Equation 2c of \cite{Bennett1962}; for the latter, note that in the proof of Lemma \ref{lem.mgf.bound}, terms in the sum in \eqref{eq.baseBennett} can be rewritten as $\{\exp(\lambda c_i)-1\}\sigma^2_i/c_i^2$. An especially low upper bound on $\sigma_i/c_i$ could reward taking this outside the sum and setting $h(u)=\exp(u)-\sum_{j=2}^\infty u^j/j!$. Over a portfolio, the probability of flooding can, and usually does, take values over the whole range between $0$ and $1$, so neither of these approaches is helpful in our application.

Similar ideas to those used in the proof of Lemma \ref{lem.mgf.bound}  can create successive improvements on a generalisation of Theorem 1 of \cite{Hoeffding1963} \cite[see][]{AnnaBarlowThesis}. However, at least in scenarios of interest to us, these did not improve upon Lemma \ref{lem.mgf.bound}, and so we do not include them in any more detail here.

\subsection{Methodology and comparison}
\label{sec.disc.meth}
When using \eqref{eq.tightestP} for simulation via the inverse-cdf method, each of the $n_{step}$ steps of a numerical solver of the two equations $F^+(s^+)=u=F^-(s^-)$  requires a numerical minimisation. This is then repeated for each of the $n_y\in\{1000,10000\}$ years and then the whole set of calculations is replicated $M\in\{100,1000\}$ times. Even though the individual minimisations are $\calO(1)$, not $\calO(n)$ (where $n$ is the size of the portfolio), Table \ref{table.CPUtime} shows that this is computationally expensive. Indeed, $n_{step}\times n_y\times M$ evaluations of Lambert's W as used in \eqref{eq.laxerP} is also only a small factor cheaper than the $\mathcal{O}(n)$ calculations used in the standard method. The sampling-importance-resampling methodology employs a tractable proposal distribution based on a well-known loosening of Bennett's inequality and reduces the number of minimisations or evaluations of Lambert's W to $n_y\times M$; Table \ref{table.CPUtime} shows the very substantial consequent gain in computational efficiency.

These simulations lead to conservative prediction intervals of the $k$-year return levels for various $k$. In our simulations, except for the two-year return level, the new intervals are between a factor of $1.5$ and $3.5$ times as wide as the intervals obtained via the standard method, with the ratio decreasing as the return level increases.

\subsection*{Acknowledgements}
Work by the first author was supported by EPSRC grant numbers EP/L015692/1 and EP/S000747/1, and JBA Risk Management. We are grateful to Paul Young and Philip Oldham of JBA Risk Management for providing advice on the nature of the sensitivity study.

\bibliographystyle{apalike}
%\bibliography{../R2AoAS/conc.bib}
\bibliography{conc.bib}

\appendix

\section{An overview of the flood-event model}\label{sec.EVTmodel}
We briefly overview the flood-event model used to provide the input to our algorithm. Specifically, each event simulated from this model leads to the following: for each insurance risk, we obtain a probability of flooding during the event and an expectation and variance of the damage ratio. These are part of the input information to the return-level simulation methodology.

\subsection{Modelling framework}

Losses due to natural disasters are commonly \citep{Kirsten2017,Franco2020,Zanardo2022} estimated using a complex modelling framework, often referred to as catastrophe (CAT) modelling \citep{Grossi2005,Foote2017,Oasis2020}. This framework brings together the simulation of extreme events with topographical knowledge and property data to simulate losses at many locations over many years of events. Our focus is on the last part of this modelling framework: Monte Carlo simulation of losses at each of a large number of locations for each extreme event over many years. For better understanding of the components in this final step in the framework we now briefly discuss earlier important parts of the framework. Definitions from this modelling framework and their corresponding notation as used in the main paper are provided in Table~\ref{tab:Defs}.

\begin{table}[h!]
\centering
\resizebox{\linewidth}{!}{%
\begin{tabular}{|l|c|l|}
	\hline 
  Term & Notation & Definition \\
  \hline
		Risk & $r$ & Depending on the detail of a portfolio, a risk \\ & & is either an insured property/location or a \\ & & collection of insured properties in a certain \\ & & postcode. \\
		Total insured value $r$ & $b_r$ & Total insured value of risk $r$. \\
		Subrisk & $s$ & Insured property/location. \\
		Portfolio & - & A set of risks and their locations. \\
%		Coverage type & $\coverage$ & Insurance coverage types (B = building, \\ & & C = contents,  BI = business interruption). \\
%		Peril type & $\peril$ & Type of flooding (river, surface water or coastal). \\ % FLRF, FLSW & STSU resp. 
%		Set of subrisks of risk $r$ & $\mathcal{S}_r$ & The collection of subrisks forming risk $r$. \\
		- & $n_r^{(\mathrm{sub})}$ & Number of subrisks in risk $r$. \\
		Flood-event model & - & Statistical extreme value model for the \\ & & simulation of flood events.\\
		& $n_y$ & Number of years in flood-event model. \\ & & Typically $n_y=1000$ or $n_y=10000$.\\
		Event set & - & Set of simulated flood events and the year in \\ & &  which they occurred. \\
%		Events in year $y$ & $\mathfrak{E}_y$ & Set of events in the year $y$ simulation. \\
		- & $n_y^{(\mathrm{ev})}$ & Number of flood events in year $y$. \\
		Flood map & - & A fine grid of simulated water depths for an \\ & &  event and peril  type covering all locations in \\ & & the portfolio, for example, the whole of the \\ & & UK. Water  depths are simulated using the \\ & & output from the flood-event model and \\ & & geographical/geological knowledge. \\
%		Hazard distribution & - & Distribution of water depths greater than 0 \\ & & at risk $r$ for event $e$. \\
		Probability of flooding & $p_{y,e,r}$ & Probability of water depth greater than 0 at \\ & & risk $r$ for event $e$. \\
		- & $\mathcal{R}_e$ & Set of risks with
a non-zero probability \\ & & of flooding in event $(y,e)$. \\
		- & $n_{y,e}^{(p>0)}$ & Number of risks with
a non-zero probability \\ & & of flooding in event $(y,e)$; $n_{y,e}^{(p>0)}=|\mathcal{R}_e|$. \\
		Damage ratio & - & Loss as a fraction of total insured value. \\
		Vulnerability function & - & Relates water depth to the mean and \\ & & standard deviation of  the damage ratio. \\ & & Depends on particular aspects of a  property \\ & & such as building type. \\
		Vulnerability distribution & - & Distribution of damage ratio for risk $r$ given \\ & & a specific water depth at the property. \\ & & Assumed to be a Beta distribution.  \\
		Wet distribution & - & Distribution of damage ratio for risk $r$ during \\ & &  event $e$.  Beta distributed with specified \\ & & parameters $\alpha_{y,e,r}$ and $\beta_{y,e,r}$ unique to this \\ & & $(y, e, r)$ combination. \\
%		Effective damage distribution & - & Distribution of damage ratio for risk $r$ due \\ $=$ loss distribution in intro? & & to flooding during event $e$. \\
%		Loss at risk $r$ in event $e$ & $L_{e,r}$ & \\
		Loss distribution & - & Distribution of loss for risk $r$ due to flooding \\ & & during event $e$. \\
Damage ratio for combination $(y,e,r)$ & $Z_{y,e,r}$ & Relative loss in event $e$ in year $y$ at risk $r$. \\
		Total loss over a portfolio in year $y$ & $T_y$ & See paper (1). \\
		Centered loss & $X_{y,e,r}$ & See paper just prior to (16). \\
		Total centered loss in year $y$ & $S_y$ & See (16).\\
Replicates of loss simulated for each year & $M$ & Typically $M=100$ or $M=1000$. \\
   \hline
\end{tabular}%
}
	\caption{Loss simulation definitions and notation.}
	\label{tab:Defs}
\end{table}

\subsubsection{Simulating flood events} \label{sec:FloodSim}

The simulation of flood events is important in understanding the flood risk and determining the loss distribution. Typically there is little loss history available and so extrapolation purely from this data would be unreliable. Thus, the mechanism that leads to these losses, \emph{i.e.}, the extreme weather events, is considered. A popular approach \citep{EA,ToweTawnLambSherlock2019,Lambetal2010,Neal2013} is to model extreme river flows at a set of points, corresponding to gauge sites, for events simulated over a long period, typically 10000 years.  Gauged river-flow data are used directly rather than rainfall data since the former incorporate spatial and temporal information, including precipitation conditions, and avoid the complexity of rainfall-runoff modelling and it’s uncertainties \citep{Lambetal2010}. The extreme river flows are modelled using (extensions of) the approach by \cite{Keef2009}, which is an extension of the Heffernan and Tawn model, a conditional approach to modelling extremes using a semi-parametric model developed by \cite{HeffernanTawn2004}, to account for temporal dependence and missing data. The spatial structure, in particular the neighbourhood used in the conditioning model, is determined using tail dependence measures. \cite{KeefTawnLamb2013} presents a practical implementation of this method using all gauges rather than the localised version used to create our event set. This spatial extreme value model is then used to simulate extreme events over a 10000 year period on a network of rainfall, river and tidal gauges. Each event is given an unique ID and the set of event IDs and the year in which they occurred is called the \textit{event set}.

Modelling spatially-conditional extremes is sensible since the occurrence and intensity of extreme values are likely to be similar at nearby locations. For example, rainfall will generally be similar at neighbouring locations as it is likely to be part of the same weather system and the topology of the area will affect the rainfall-runoff process. Under the \cite{Keef2009} approach covariates are not explicitly incorporated into the extreme value distribution parameters since dependence is already embedded into the procedure directly through conditioning on the extremes of neighbouring river flow processes \citep{Gilleland2013}. Nevertheless inclusion of covariates such as rainfall may also be considered at this stage; in particular \cite{Jonathan2013} extend the Heffernan and Tawn model to incorporate covariate effects in threshold selection; see also \cite{WinterTawnBrown2016}. 

\subsubsection{Water depths and distribution}

In the second step of the CAT modelling framework the simulated extreme events from step 1 are used to drive an hydraulic model to obtain water depths (essentially) everywhere, at a 5/30m resolution over the UK/globally using JBA's software JFLOW \citep{JBAREF}. One implementation of this stage is detailed in \cite{Neal2013} and an overview of flood inundation modelling approaches is given in \cite{Teng2017}. To create the map of water depths many covariates are considered from land-use and soil water retention to elevation. Flood maps have also been created to incorporate (possible future) changes due to climate change \citep{JBAREFtwo}. 

\subsubsection{Portfolio information and vulnerability}

The portfolio provides an ID and location for each risk in terms of postcode and sometimes also latitude and longitude. Some risks may in fact be a collection of multiple insured properties, referred to as \textit{subrisks}, all assigned to the same location and ID. There is no information on individual subrisks only the whole collection, such as the total value of the subrisks, denoted by $b_r$, and the number of subrisks. For each subrisk, $s$, there is a mapping to the risk it is part of: $r=r(s)$. The number of subrisks of risk $r$ is $n_r^{(\mathrm{sub})}$.

In order to translate the water depths into a loss some measure of how different water depths affect the resulting loss is needed. The \textit{vulnerability function} gives a mean and standard deviation of the relative loss (fraction of the value of a risk that will be lost, termed the \textit{damage ratio}) for each of a discrete number of depths. Each risk will have a different set of expectations and variances of the relative loss due to particular aspects of the property. The \textit{vulnerability distribution} of the damage caused relative to the value of the risk given a certain water depth is assumed to follow a Beta distribution with parameters such that the expectation and standard deviation match those given by the vulnerability function.

\subsubsection{Damage distribution and loss}

The modelled water depths are then combined with the vulnerability to create a loss distribution for each event at each risk in the portfolio. The damage ratio at a particular risk for a particular event is 0 with some probability $1-p_{y,e,r}$, and follows a \textit{wet distribution} otherwise. This wet distribution is a mixture of Beta distributions (vulnerability) where the weights are determined by the distribution of water depths, and it is approximated by a Beta distribution. The loss at a particular risk for an event is then the sum over the subrisks of the damage ratios multiplied by the average total insured value of a subrisk.

Given the water depth distribution (which incorporates topological information about the risk location and the extreme event modelling) and the vulnerability (individual assessment of possible damagee at each risk), the individual losses are considered independent. All relevant covariates are assumed to have been fed through the previous stages of the CAT model framework, therefore they are not used in the final modelling stage.

\subsection{Applications and impact}

The CAT modelling procedure is employed in many areas of insurance and natural hazard risk management \citep{Mahdyiar2005}; for example, for earthquake \citep{Bommer2002,Akkar2021}, wildfire \citep{Murnane2006,Oliveira2021} and hurricane \citep{Pinelli2008} loss calculations. The framework allows one to not only assess the exposure of a current portfolio but also to perform stress tests and 'what if?' analyses. Losses also do not necessarily need to be in terms of money. Human and environmental impact could also form part of a loss function, similar to the vulnerability function. The resulting assessments of total loss improve public authorities understanding of risk and can help to inform decisions on, for example, flood management and land development. The public sector can also benefit from this loss modelling to assess and address the insurance protection gap, especially in the context of climate change \cite{EIOPArefONE,EIOPArefTWO}.

\section{Proofs of Lemma \ref{lem.mgf.bound} and Theorem \ref{thrm.new}}
  \label{sec.proofs}
This section proves Theorem \ref{thrm.new}. We first note some properties of the function $f_2$; from the first of these we prove \eqref{eq.new.mgf.bound.tightest} and bound the discrepancy from a tightest-possible Bennett-like bound, giving a lower bound on the best possible Bennett-like concentration inequality that accounts for the individual $c_i$.  From \eqref{eq.new.mgf.bound.tightest} and Proposition \ref{lemma.f.properties}(ii) we derive our new concentration inequalities.

\begin{proposition}
  \label{lemma.f.properties}
(i) The Maclaurin series for $f_2(u)$ is $\sum_{j=0}^{\infty}u^j/(j+2)!$, and (ii) 
      $f_2'(u)=(1-2/u)f_2(u)+1/u$.
\end{proposition}

\subsection{Proof of Lemma \ref{lem.mgf.bound}}
  \begin{proof}
As in the proof of Bennett's inequality, for any random variable $X$ with $\Expect{X}=0$ and $\Prob{X\le c}=1$,
$$
\begin{aligned}
  \Expect{\exp(\lambda X)}
  &=
  \Expect{\sum_{j=0}^\infty\frac{\lambda^j}{j!} X^j}
  =
  \sum_{j=0}^\infty\frac{\lambda^j}{j!} \Expect{X^j}
  =
  1+\sum_{j=2}^\infty\frac{\lambda^j}{j!} \Expect{X^j}\\
  &\le
  1+ \sum_{j=2}^\infty\frac{\lambda^j}{j!} \Expect{X^2}c^{j-2}
  =
  1+\lambda^2 \mathsf{Var}[X]f_2(\lambda c)\\
  &\le \exp\{\lambda^2 \mathsf{Var}[X]f_2(\lambda c)\}.
  \end{aligned}
$$

Again, as in the proof of Bennett's inequality, since the $X_i$ are independent,
\begin{equation}
  \label{eq.baseBennett}
\frac{1}{n}\log\mathbb{E}[\exp(\lambda S_n)]=\frac{1}{n}\sum_{i=1}^n \log \Expect{\exp(\lambda X_i)}
\le
\lambda^2\frac{1}{n}\sum_{i=1}^n\sigma^2_if_2(\lambda c_i).
\end{equation}
Now, let
$$
h(u)=\frac{1}{2!}+\frac{u}{3!}=f_2(u)-\sum_{j=2}^\infty \frac{u^{j}}{(j+2)!},
$$

by Proposition \ref{lemma.f.properties}. Rearranging the above also shows that the infinite sum is absolutely convergent for all finite $u$. Since $h$ is linear,  $h(u)= (1-u/u_*)h(0)+(u/u_*)h(u_*)$. Rearranging, and then considering only $0\le u \le u_*$, we obtain:
\begin{align}
  f_2(u)&=\left(1-\frac{u}{u_*}\right)f_2(0)+\frac{u}{u_*}f_2(u_*)
  +\sum_{j=2}^\infty \frac{u^j}{(j+2)!}-\frac{u}{u_*}\sum_{j=2}^\infty \frac{u_*^j}{(j+2)!} \nonumber \\
   \label{eq.bound.all.j}
  &=\left(1-\frac{u}{u_*}\right)f_2(0)+\frac{u}{u_*}f_2(u_*)
  +u^2\sum_{j=0}^\infty \frac{u^j}{(j+4)!}
  -uu_*\sum_{j=0}^\infty \frac{u_*^j}{(j+4)!}\\
  &\le\left(1-\frac{u}{u_*}\right)f_2(0)+\frac{u}{u_*}f_2(u_*)
  +u^2\sum_{j=0}^\infty \frac{u_*^j}{(j+4)!}-uu_*\sum_{j=0}^\infty \frac{u_*^j}{(j+4)!}\nonumber \\
%  &=
%\left(1-\frac{u}{u_*}\right)f_2(0)+\frac{u}{u_*}f_2(u_*)-\frac{u}{u_*}\left\{1-\frac{u}{u_*}\right\}\sum_{j=2}^\infty\frac{u_*^j}{(j+2)!}\\
%  &=
%\left(1-\frac{u}{u_*}\right)f_2(0)+\frac{u}{u_*}f_2(u_*)-\frac{u}{u_*}\sum_{j=2}^\infty\frac{u_*^j}{(j+2)!}\left\{1-\frac{u^{j-1}}{u_*^{j-1}}\right\}\\
%\nonumber
%&=
%\left(1-\frac{u}{u_*}\right)f_2(0)+\frac{u}{u_*}f_2(u_*)-\frac{u}{u_*}\left\{1-\frac{u}{u_*}\right\}\sum_{j=2}^\infty\frac{u_*^j}{(j+2)!}\\
&=
\label{eq.individ.bound}
f_2(0)+\frac{u}{u_*}\left\{f_2(u_*)-f_2(0)\right\}-\frac{u}{u_*}\left\{1-\frac{u}{u_*}\right\}u_*^2 f_4(u_*),
\end{align}
where the inequality follows because $0\le u\le u_*$.
Thus
\begin{align*}
\sum_{i=1}^n \sigma_i^2f_2(\lambda c_i)
&\le
\sum_{i=1}^n
\left\{\sigma_i^2 f_2(0)+\sigma_i^2\frac{c_i}{c_*}\left\{f_2(\lambda c_*)-f_2(0)\right\}-\sigma_i^2\frac{c_i}{c_*}\left(1-\frac{c_i}{c_*}\right)\lambda^2c_*^2f_4(\lambda c_*)\right\}\\
&=
n\left\{\sigmasqbar f_2(0)+K \{f_2(\lambda c_*)-f_2(0)\}
-
\lambda^2c_*^2 K_1 f_4(\lambda c_*)\right\},
\end{align*}
which simplifies to \eqref{eq.new.mgf.bound.tightest} on multiplying by $\lambda^2/n$, as $f_2(0)=1/2$.

Next, the discrepancy between the bound we use, \eqref{eq.individ.bound}, and $f_2(u)$ in \eqref{eq.bound.all.j}  is
\[
u^2\sum_{j=0}^\infty\frac{u_*^j}{(j+4)!}\left\{1-\frac{u^{j}}{u_*^{j}}\right\}
=
u^2\sum_{j=1}^\infty\frac{u_*^j}{(j+4)!}\left\{1-\frac{u^{j}}{u_*^{j}}\right\}
\le
u^2u_*\sum_{j=0}^\infty\frac{u_*^j}{(j+5)!}=\frac{u^2}{u_*^2}u_*^3f_5(u_*).
\]
Hence
\[
B(\lambda; c_{*},\sigmasqbar,K,K_1)-\frac{1}{n}\lambda^2\sum_{i=1}^n \sigma_i^2f_2(\lambda c_i)
\le
\frac{1}{n}\lambda^2\sum_{i=1}^n \sigma_i^2\frac{c_i^2}{c_*^2}(\lambda c_*)^3f_5(\lambda c_*)
=
\lambda^5c_*^3(K-K_1)f_5(\lambda c_*).
\]
  \end{proof}

\subsection{Proof of Theorem \ref{thrm.new}}

  \begin{remark}
    \label{remark.summary.order}
    Since $0\le c_i\le c_*$, we have $K_1\le K\le \sigmasqbar$.
  \end{remark}

  We now prove Theorem \ref{thrm.new}.
  
  \begin{proof}
    The tightest bound, \eqref{eq.tightestP}, follows directly from \eqref{eq.Chernhoff} and \eqref{eq.new.mgf.bound.tightest}. Since $K_1>0$, setting $K_1=0$ gives the looser bound \eqref{eq.looser.problem}, which we solve to obtain $\lambda_*$, leading directly to \eqref{eq.laxerP} and \eqref{eq.laxestP}. Taking \eqref{eq.new.mgf.bound.tightest} with $K_1=0$ and using Proposition \ref{lemma.f.properties} (ii),
    \begin{align*}
    B'(\lambda; c_{*},\sigmasqbar,K,0)
&    =
    \lambda \sigmasqbar +2\lambda K\left\{f_2(\lambda c_*)-\frac{1}{2}\right\}+\lambda^2K c_*f'(\lambda c_*)\\
&    =
    \lambda \sigmasqbar +2\lambda K\left\{f_2(\lambda c_*)-\frac{1}{2}\right\}+\lambda^2K c_*\left\{\left(1-\frac{2}{\lambda c_*}\right)f_2(\lambda c_*)+\frac{1}{\lambda c_*}\right\}\\
    &=
    \lambda \sigmasqbar + \lambda^2K c_* f_2(\lambda c_*)
    =\lambda \sigmasqbar + \frac{K}{c_*}\left\{\exp(\lambda c_*)-1-\lambda c_*\right\}\\
   & =\lambda \left(\sigmasqbar-K\right)-\frac{K}{c_*}+\frac{K}{c_*}\exp(\lambda c_*)
    \end{align*}
Thus, to minimise $B(\lambda)-\lambda t$ we must solve
    \[
\frac{K}{c_*}\exp(\lambda c_*) =
    -\lambda \left(\sigmasqbar-K\right)+\left(t+\frac{K}{c_*}\right).    
    \]
    Since the left hand side is increasing, with an intercept at $\lambda=0$ which is strictly less than the intercept of the right hand side, and the right hand side is decreasing (see Remark \ref{remark.summary.order}), there is exactly one solution on $[0,\infty)$. We must solve
    \begin{align*}
      K\exp(\lambda c_*)= (K+tc_*)-\lambda c_*(\sigmasqbar -K)
      ~~\iff
      \frac{K}{\sigmasqbar-K}=\left[\frac{K+tc_*}{\sigmasqbar-K}-\lambda c_*\right]\exp(-\lambda c_*)\\
     ~~~\iff
      \frac{K}{\sigmasqbar-K}\exp\left(\frac{K+tc_*}{\sigmasqbar-K}\right)
      =
      \left[\frac{K+tc_*}{\sigmasqbar-K}-\lambda c_*\right]
      \exp\left(\frac{K+tc_*}{\sigmasqbar-K}-\lambda c_*\right).      
\end{align*}
    Thus
    \[
    \frac{K+tc_*}{\sigmasqbar-K}-\lambda c_*
    =
    \calW\left(\frac{K}{\sigmasqbar-K}\exp\left(\frac{K+tc_*}{\sigmasqbar-K}\right)\right),
    \]
    which leads to \eqref{eq.lamstar}.
    \end{proof}

\section{Further tightening of the bound \eqref{eq.new.mgf.bound.tightest}}
\label{app.tighter.still}
As alluded to in the discussion, \eqref{eq.new.mgf.bound.tightest} may be tightened still further provided additional summaries are created.

When simplifying \eqref{eq.bound.all.j}, for a fixed $J\ge 1$,

\begin{align*}
  u^2\sum_{j=0}^\infty\frac{u^j}{(j+4)!}
&=
%u^2\sum_{j=0}^{J-1}\frac{u^j}{(j+4)!}+u^2\sum_{j=J}^\infty \frac{u^j}{(j+4)!}
%=
u^2\sum_{j=0}^{J-1}\frac{u^j}{(j+4)!}+u^{2+J}\sum_{j=0}^\infty \frac{u^j}{(j+4+J)!}\\
&\le
u^2\sum_{j=0}^{J-1}\frac{u^j}{(j+4)!}+u^{2+J}\sum_{j=0}^\infty \frac{u_*^j}{(j+4+J)!}
=
u^2\sum_{j=0}^{J-1}\frac{u^j}{(j+4)!}+u^{2+J}f_{4+J}(u_*).\end{align*}
Similarly
\[
  uu_*\sum_{j=0}^\infty\frac{u_*^j}{(j+4)!}=uu_*\sum_{j=0}^{J-1}\frac{u_*^j}{(j+4)!}+uu_*^{1+J}f_{4+J}(u_*).
\]
For $J\ge 1$, this leads to the tighter limit of
\begin{align*}
  f_2(u)
  &\le
  f_2(0)+\frac{u}{u_*}\left\{f_2(u_*)-f_2(0)\right\}
  -\frac{u}{u_*}\sum_{j=0}^{J-1}\frac{u_*^{j+2}}{(j+4)!}\left\{1-\frac{u^{j+1}}{u_*^{j+1}}\right\}
  - \frac{u}{u_*}\left\{1-\frac{u^{J+1}}{u_*^{J+1}}\right\} u_*^{J+2} f_{J+4}(u_*).
\end{align*}
The upper bound \eqref{eq.new.mgf.bound.tightest} for $n^{-1}\log \Expect{\exp(\lambda S_n)}$ is then replaced with
\[
    \frac{1}{2}\lambda^2 \sigmasqbar+\lambda^2 K\left\{f_2(\lambda c_*)-\frac{1}{2}\right\}
    -
    \sum_{j=0}^{J-1}\frac{K_{j+1}}{(j+4)!}\lambda^{j+4}c_*^{j+2}
-K_{J+1}\lambda^{J+4}c_*^{J+2} f_{J+4}(\lambda c_*),
\]
where $K_j=n^{-1}\sum_{i=1}^n \sigma^2_i \frac{c_i}{c_*}\left(1-\frac{c_i^j}{c_*^j}\right)$ generalises the definition of $K_1$.

\section{Additional Figures}
\label{sec.addfigs}

\begin{figure}
\begin{center}
  \includegraphics[scale=0.39,angle=0]{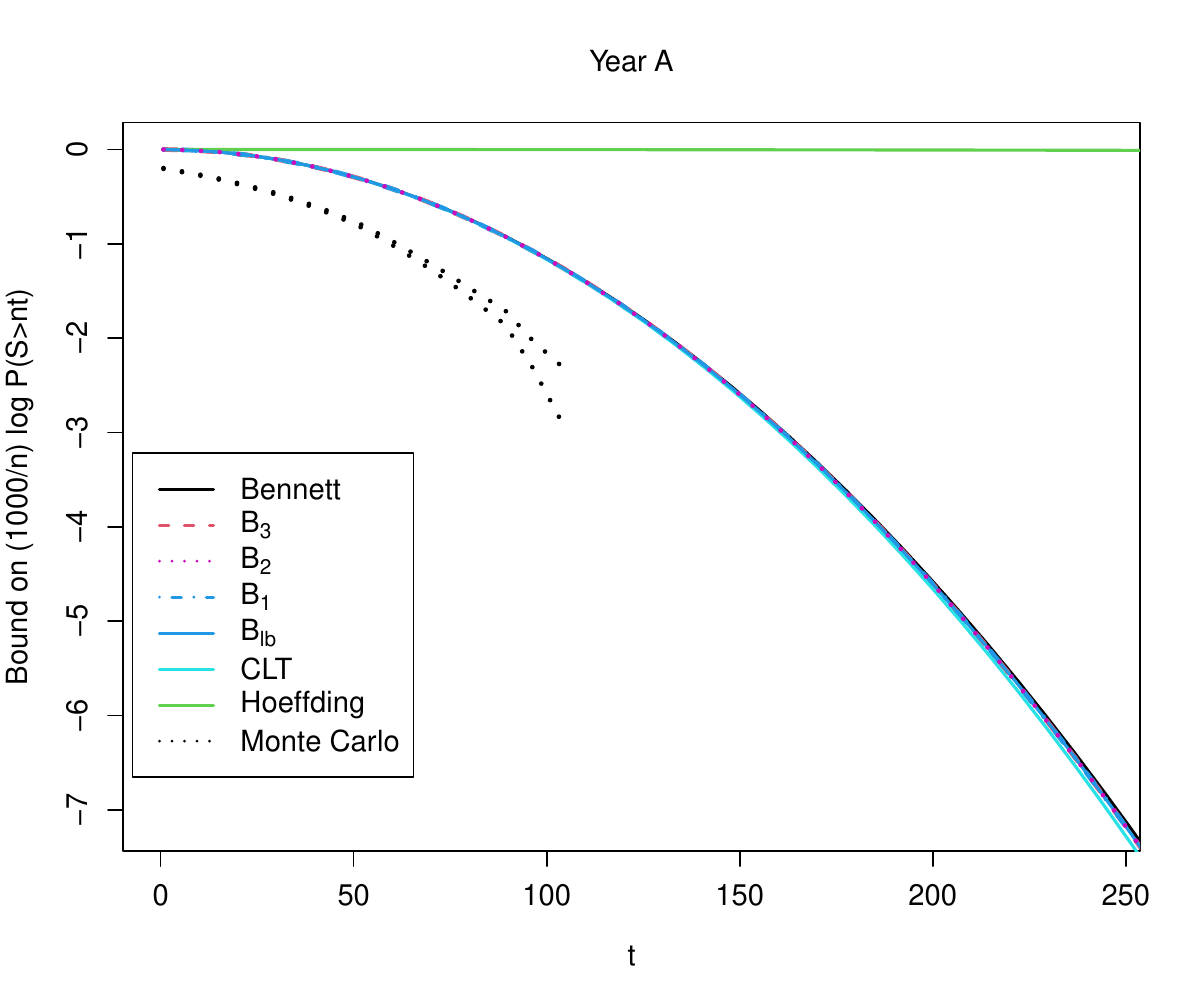}
  \includegraphics[scale=0.39,angle=0]{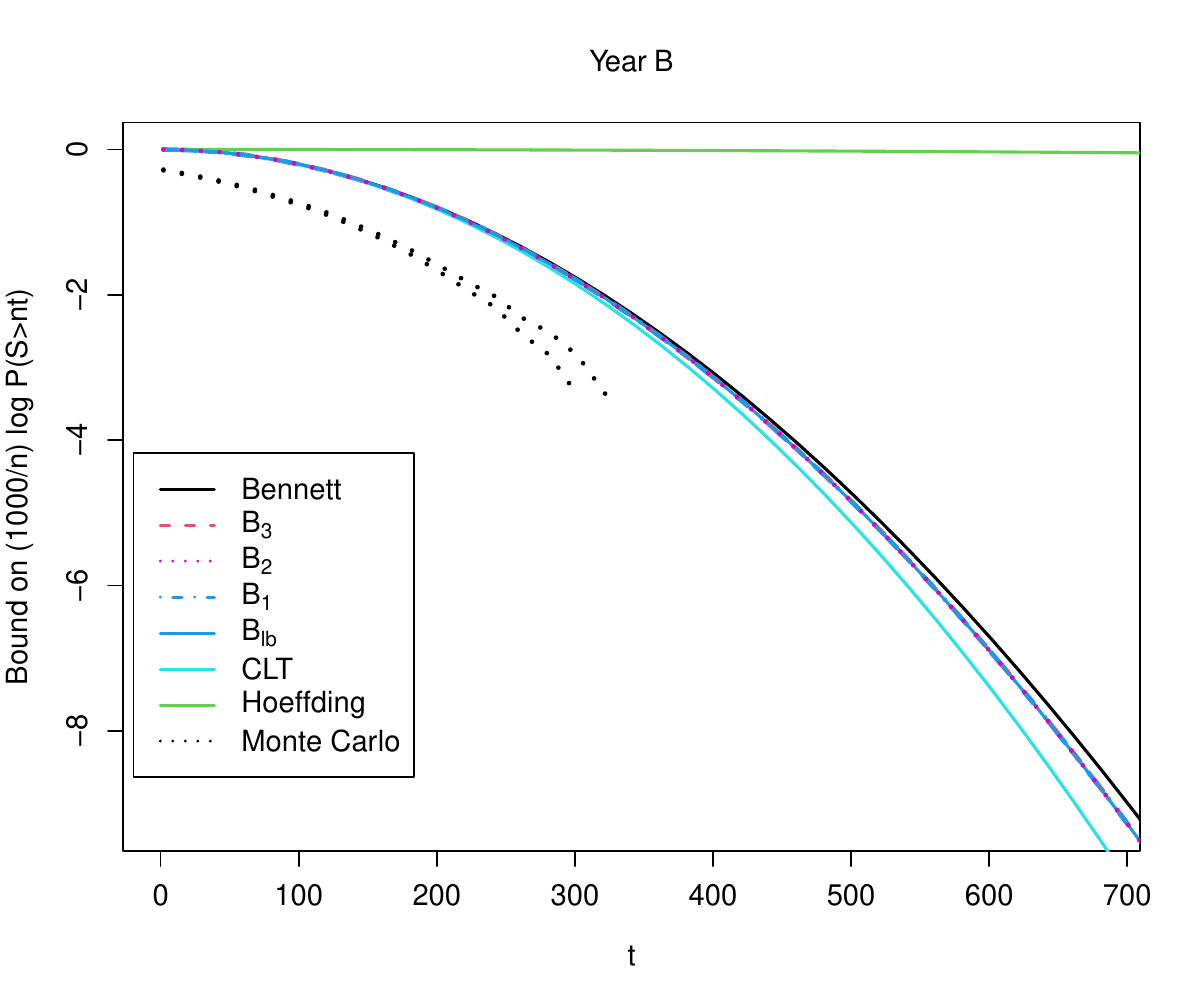}
  \includegraphics[scale=0.39,angle=0]{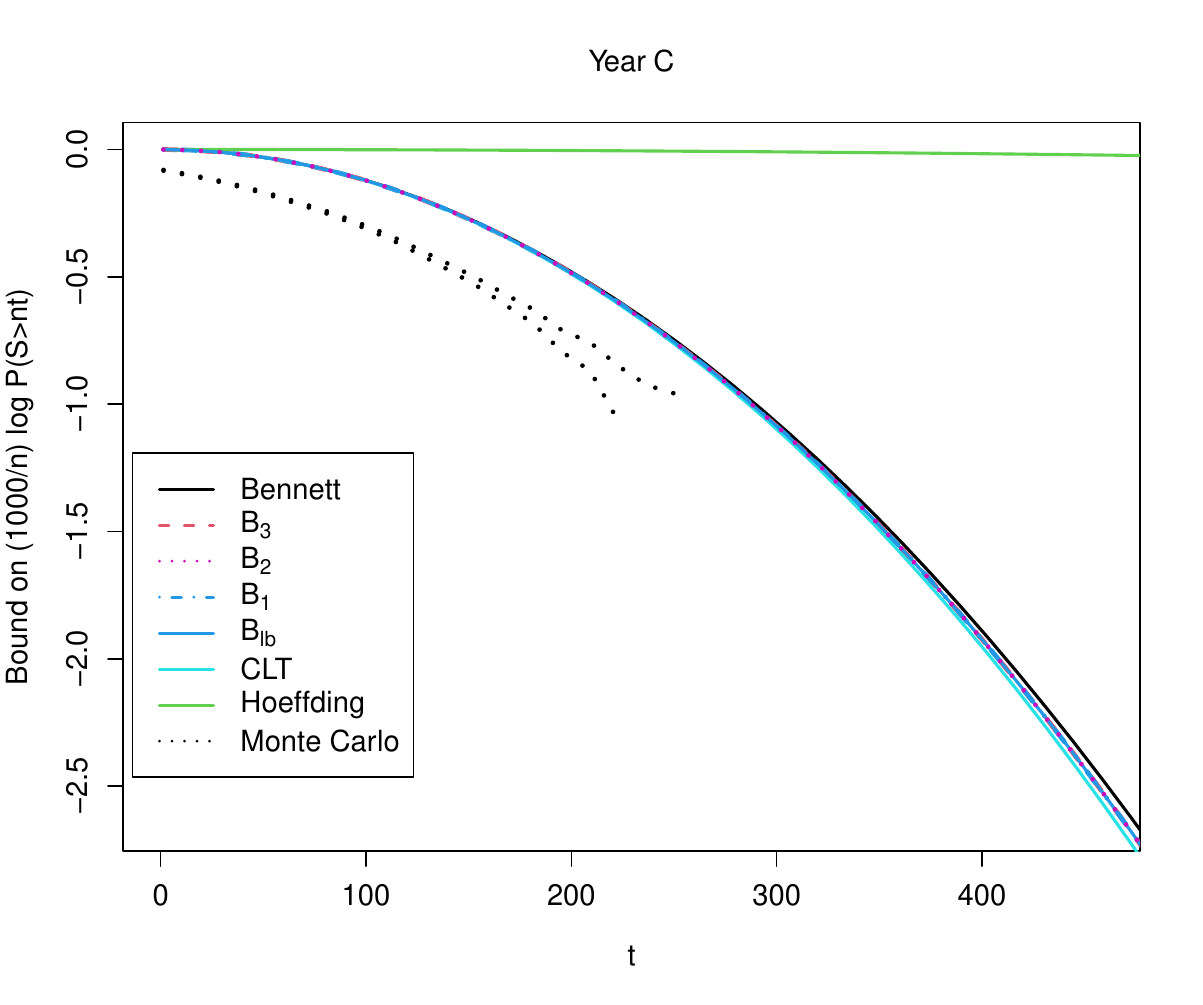}
  \includegraphics[scale=0.39,angle=0]{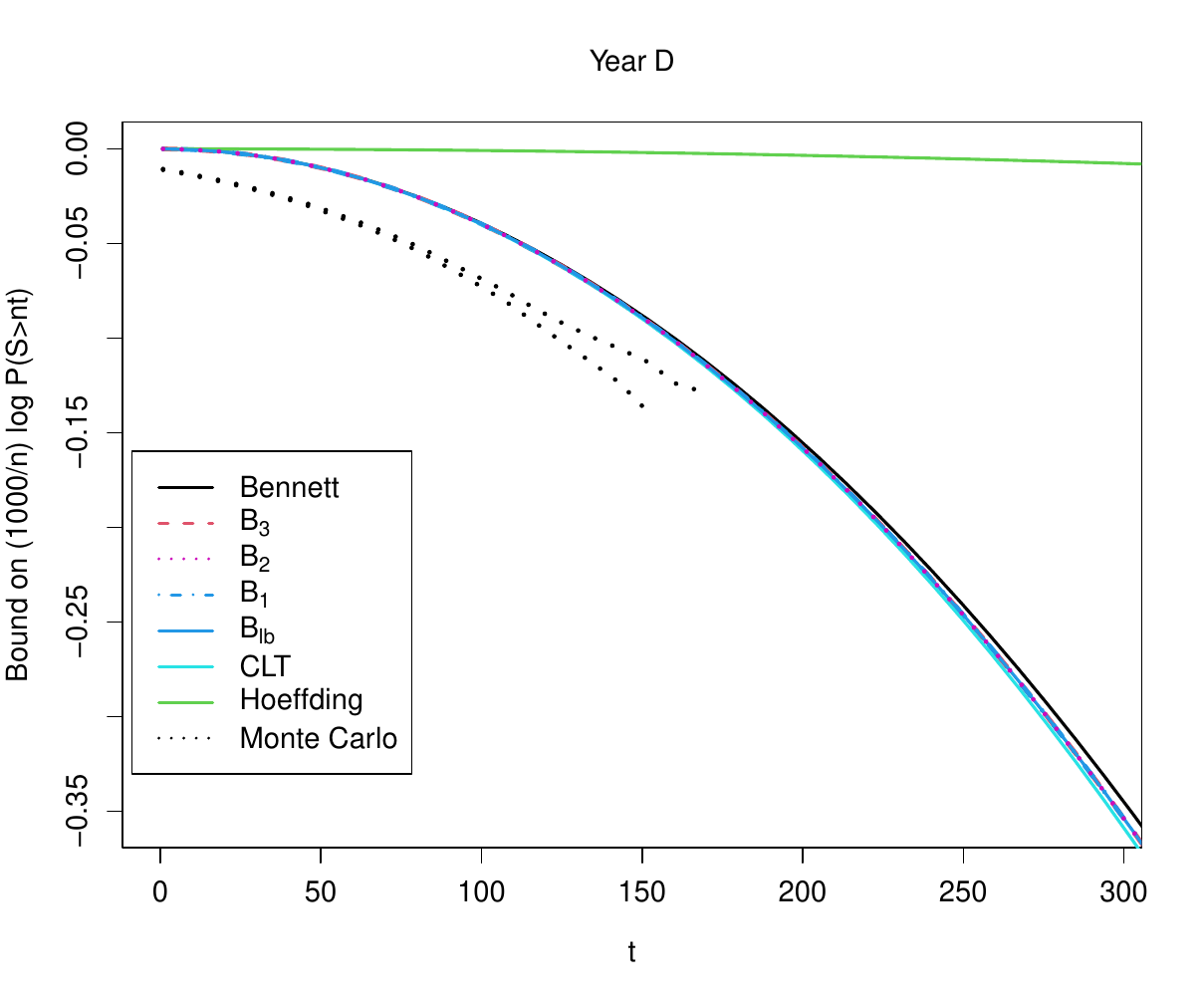}
  \caption{Comparison of bounds on $(1000/n)\log \Prob{S_n\le -nt}$ plotted against $t$. The Bennett bound is shown as a black solid line; Equation \eqref{eq.tightestP}, denoted $B_1$, is in blue dot-dash and Equation \eqref{eq.laxestP}, denoted $B_3$, is in red dashed. Other solid lines are: the large-$n$ approximation from the central limit theorem (cyan) and the Hoeffding bound (green). The dotted lines show 90\% CIs of the true probability based on $20000$ Monte Carlo simulations from the true distribution. The four panels correspond to centred random variables, respectively from Years A, B, C and D from Section \ref{sec.stdPortfolio}.
\label{fig.data.FourYearsrev}
}
\end{center}
\end{figure}

Figure \ref{fig.data.FourYearsrev} is the converse of Figure \ref{fig.data.FourYears}, estimating $n^{-1}\Prob{S_n-\mathbb{E}[S_n]\le -nt}$ rather than\\
$n^{-1}\Prob{S_n-\mathbb{E}[S_n]\ge nt}$. The results for the lower tail are so different to those for the upper tail for the years with lower expected total cost because the distributions of the total cost are skewed. For example, the (positive) skewness for the four years A, B, C and D, estimated from 20000 simulations, are approximately $1.20$ and $0.47$, $0.18$ and $0.08$ respectively.

\newpage
\section{Derivation of $\md \lambda /\md t$}
\label{eqn.sec.derive.dlambydt}
From \eqref{eq.lamstar}, 
\[
\lambda = \frac{1}{c_*}\left[r(t)-\calW\left(\frac{nK}{c_*}r'(t)\exp[r(t)]\right)\right]
~~~\mbox{where}~~~
r(t)=\frac{K+t c_*/n}{\sigmasqbar-K}.
\]
Since $r''(t)=0$,
\[
\frac{\md\lambda}{\md t}=\frac{1}{c_*}\left[r'(t)-\frac{nK}{c_*}\{r'(t)\}^2\exp[r(t)]\calW'\left(\frac{nK}{c_*}r'(t)\exp[r(t)]\right)\right]
=
\frac{r'(t)}{c_*}\left[1-x\calW'\left(x\right)\right],
\]
where $x=\frac{nK}{c_*}r'(t)\exp[r(t)]$.

But, by definition, $x=W\exp(W)$, where $W=\calW(x)$, so $dx/dW=(W+1)\exp(W)=(W+1)x/W$, so $xW'(x)=W/(1+W)$. Thus
\[
\frac{\md\lambda}{\md t}=\frac{r'(t)}{c_*}\frac{1}{1+W}=\frac{1}{n(\sigmasqbar-K)}\times \frac{1}{1+W},
\]
where $W$ is as defined in \eqref{eq.W.for.dlambydt}.

\section{Comparison of importance sampling-resampling with direct simulation}
\label{sec.cmpISwithDirect}
\begin{figure}
\begin{center}
  \includegraphics[scale=0.85,angle=0]{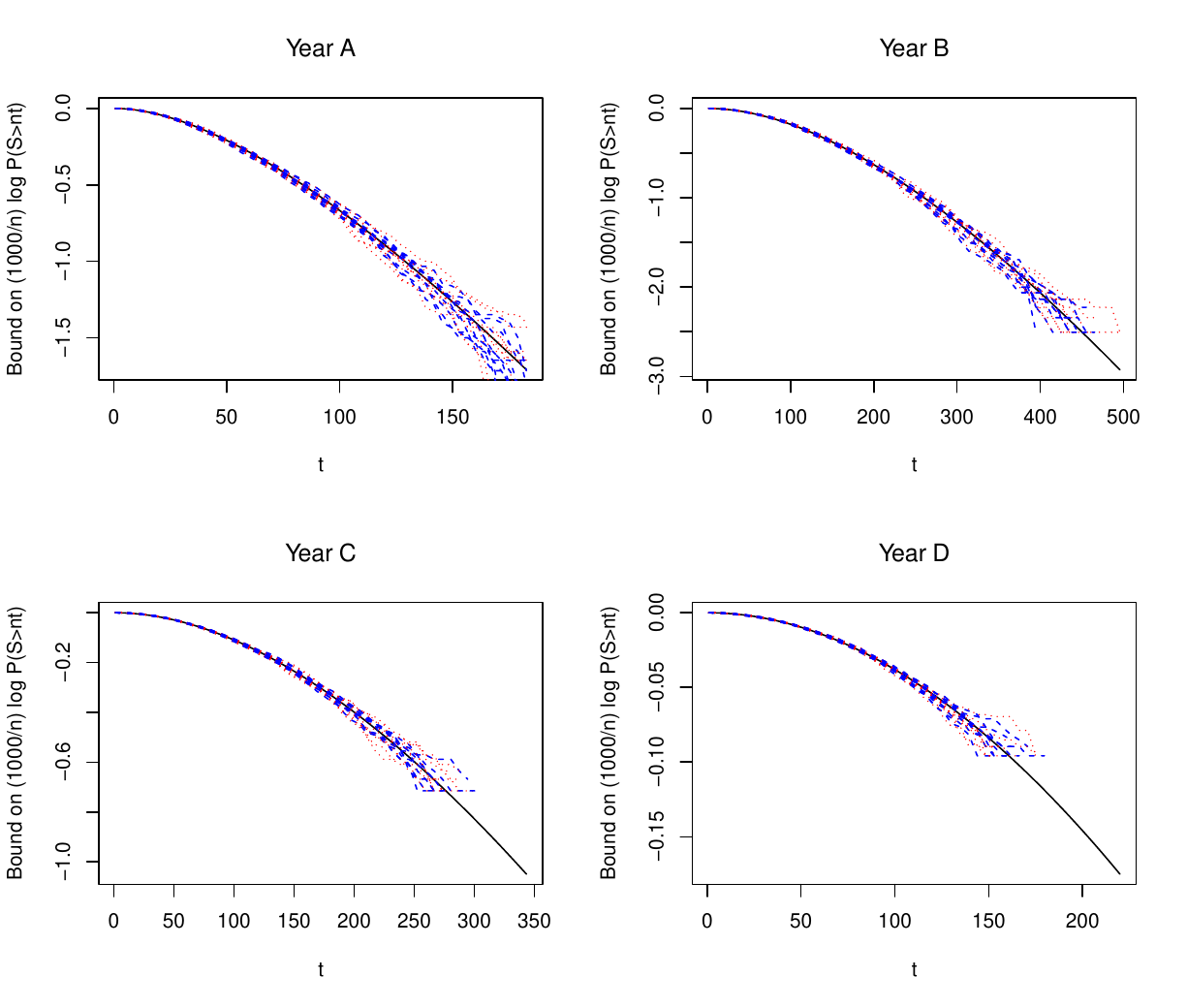}
  
  \caption{Comparison of the bound $B_2$ (black) and empirical estimates of the bound by ten replicates of $1000$ simulations from the loss for that year via the $B_2$ bound, both via direct sampling (shown in blue) and via sampling-importance-resampling (shown in red). The lines cut off when the empirical estimate of the probability is $1/1000$. The four panels correspond to the four representative years, A, B, C and D from Section \ref{sec.stdPortfolio}.
    % sometimes its 2/1000 then the next t in the sequence gives 0/1000
    \label{fig.logProb.directvSIR}
}
\end{center}
\end{figure}

For each of the four representative years from Section \ref{sec.stdPortfolio}, Figure \ref{fig.logProb.directvSIR} compares the true bound, $B_2$, against estimates of it obtained via Monte Carlo simulations using the direct method and using sampling-importance-resampling. It suggests that for each year, each simulation method leads to similar variability about the true probability under $B_2$. 

\begin{figure}
\begin{center}
  \includegraphics[scale=0.8,angle=0]{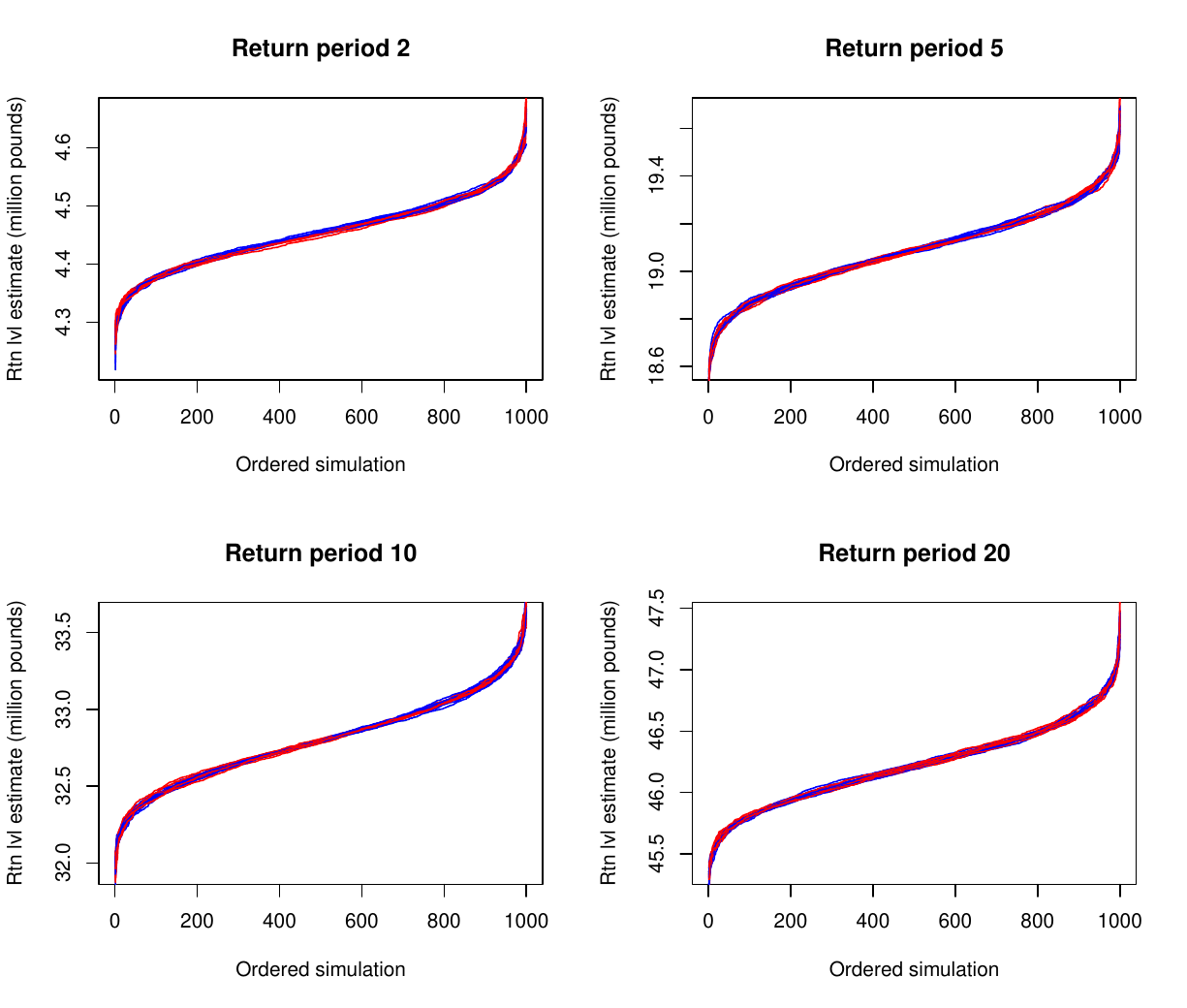}

  \caption{Comparison of simulations of the upper bound on the $k$-year return level for $k\in\{2,5,10,20\}$. All simulations use $B_2$ and $F^+_y$. Results from direct sampling are shown in blue, whilst results from importance sampling-resampling appear in red. Each plot corresponds to a particular return period and shows ten replicate lines for each bound. Each line is derived from $M=1000$ simulations, ordered by the simulated value.  
\label{fig.rtnlvl.directvSIRa}
}
\end{center}
\end{figure}

\begin{figure}
\begin{center}
  \includegraphics[scale=0.8,angle=0]{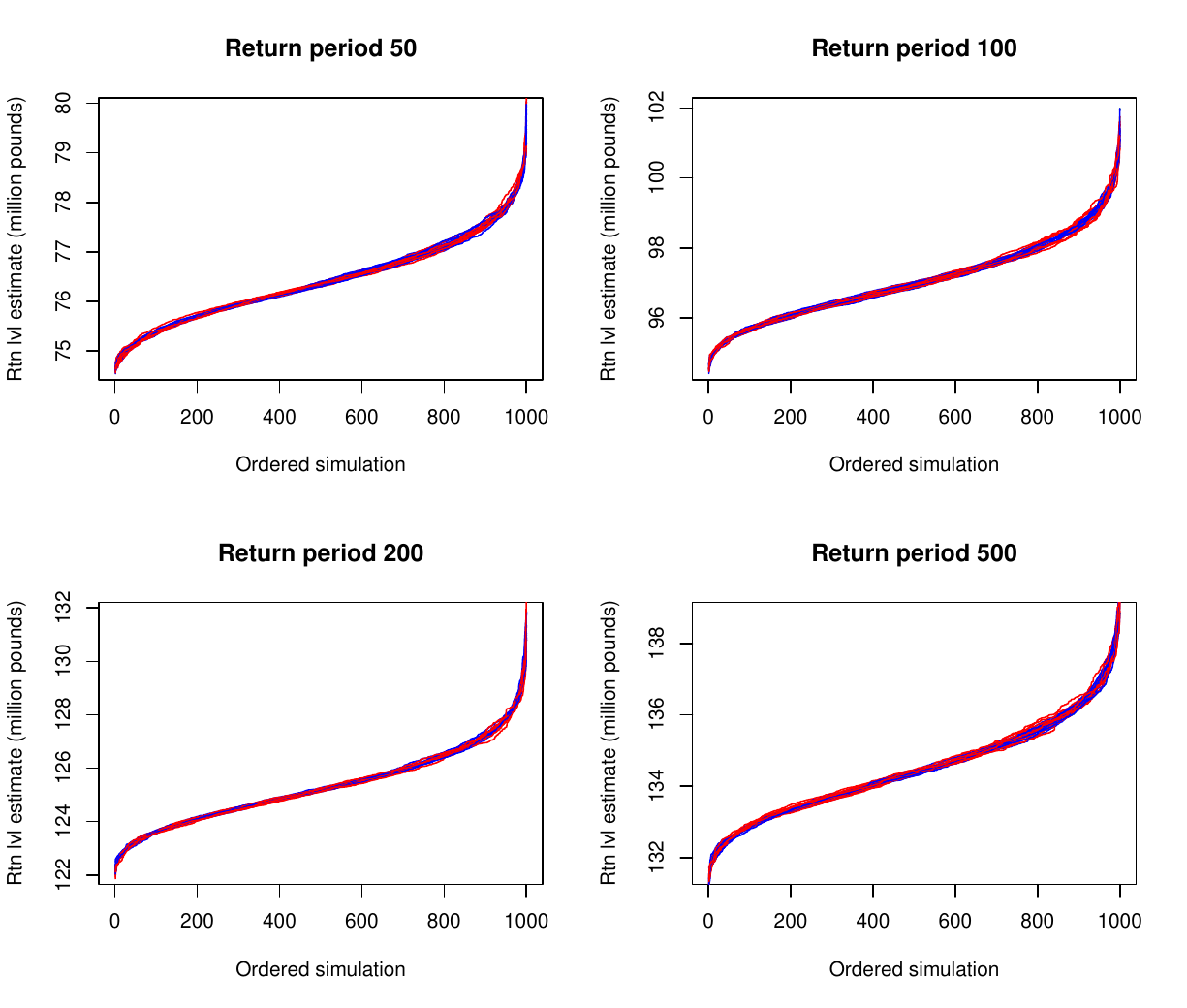}

  \caption{Comparison of simulations of the upper bound on the $k$-year return level for $k\in\{50,100,200,1000\}$. All simulations use $B_2$ and $F^+_y$. Results from direct sampling are shown in blue, whilst results from importance sampling-resampling appear in red. Each plot corresponds to a particular return period and shows ten replicate lines for each bound. Each line is derived from $M=1000$ simulations, ordered by the simulated value.  
\label{fig.rtnlvl.directvSIRb}
}
\end{center}
\end{figure}

Figures \ref{fig.rtnlvl.directvSIRa} and \ref{fig.rtnlvl.directvSIRb} compare the return-level distributions obtained from simulations from all $1000$ years using direct simulation and then using sampling-importance-resampling, and suggests that any discrepancies between the distributions are small and, except for the two-year return levels,  (at most) of a similar magnitude to the Monte Carlo variability between replicates.

\section{Further return-level estimates}
\label{app.more.rtn.lvls}
\begin{figure}
\begin{center}
  \includegraphics[scale=0.39,angle=0]{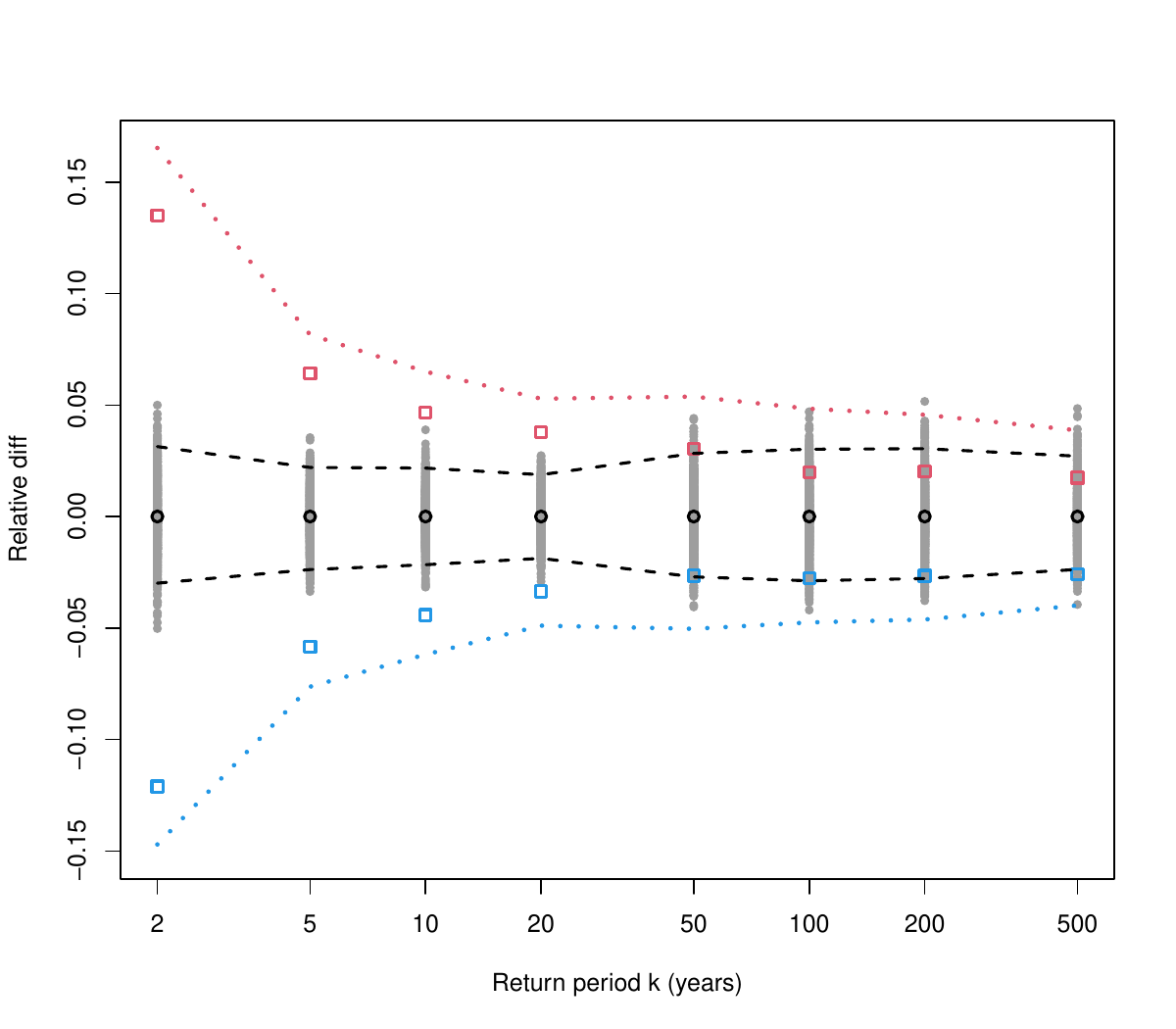}
  \includegraphics[scale=0.39,angle=0]{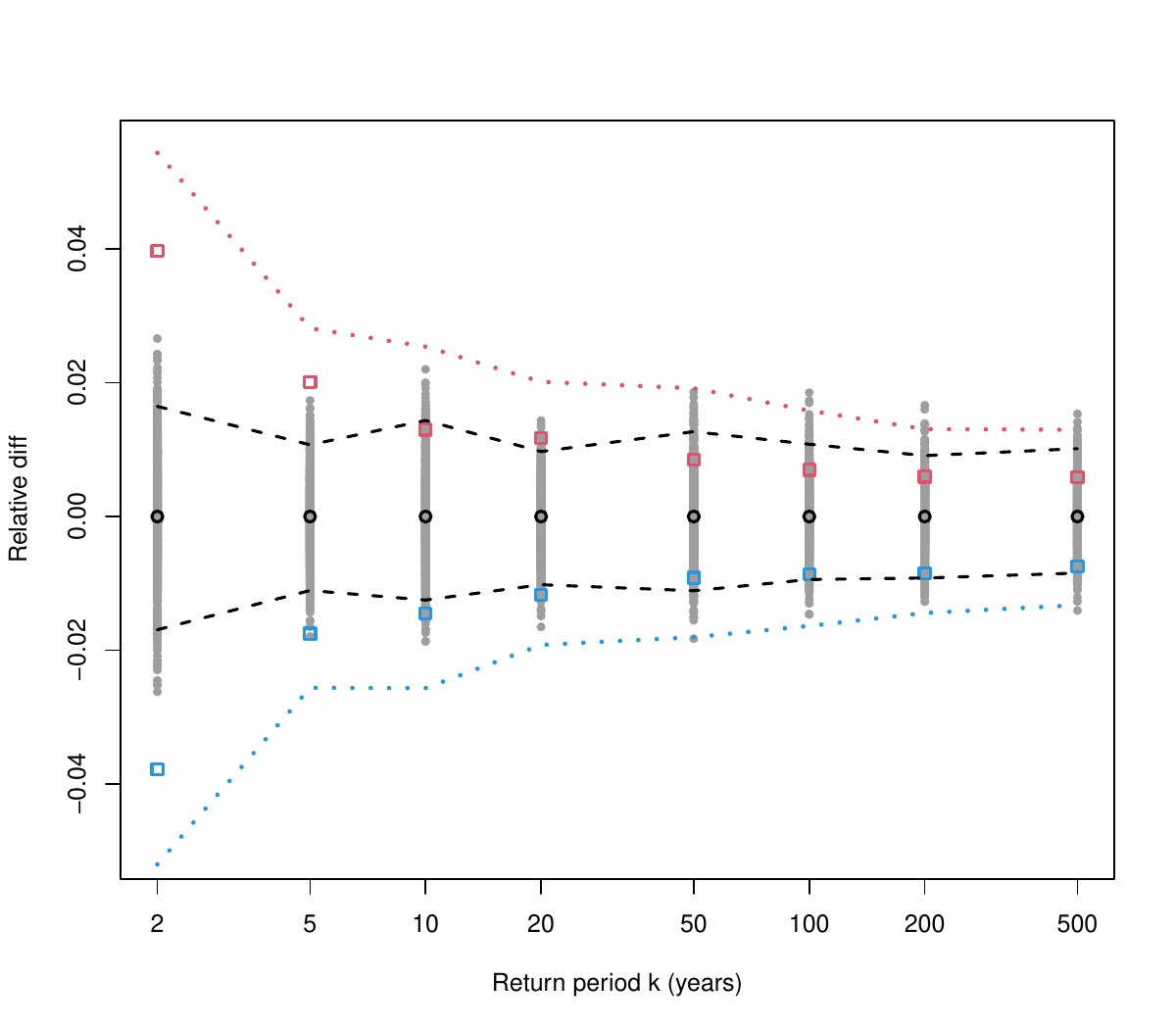}
  \caption{Relative return level point estimates and 95\% predictive bounds via the standard method (in black) and via sampling importance resampling using bound $B_2$ for $F^+$ and $F^-$. The point estimate and upper end of the prediction interval when using $F^+$ appear in red; when using $B_2$ for $F^-$, the corresponding point estimate and lower predictive bound appear in blue. Values are relative to the point estimate from the standard method, $s$; \emph{i.e.}, $v:\to (v-s)/s$. The first is for the standard portfolio using a different random seed to that which generated the results in Figure \ref{fig.rtnlvlcomp}. The second plot is for the larger portfolio. In both plots, the $1000$ realisations from the standard method appear in grey.
\label{fig.rtnlvlcompB}
}
\end{center}
\end{figure}

Figure \ref{fig.rtnlvlcompB} repeats the second plot in Figure \ref{fig.rtnlvlcomp} but with a different random seed, and shows the same overall pattern. It then repeats this plot again but for the large portfolio. The overall pattern is the same as for the standard portfolio, but we highlight two changes. Firstly, the relative differences are smaller, as would be expected since yearly expected losses are proportional to $n$, the portfolio size, whereas standard deviations are proportional to $\sqrt{n}$. Furthermore, the disrepancies between the predicted CIs using $B_2$ and those using the standard method are smaller.

\section{Sensitivity study for 20-year return level}\label{app:sensitivity}

\begin{figure}
\begin{center}
\includegraphics[scale=0.6]{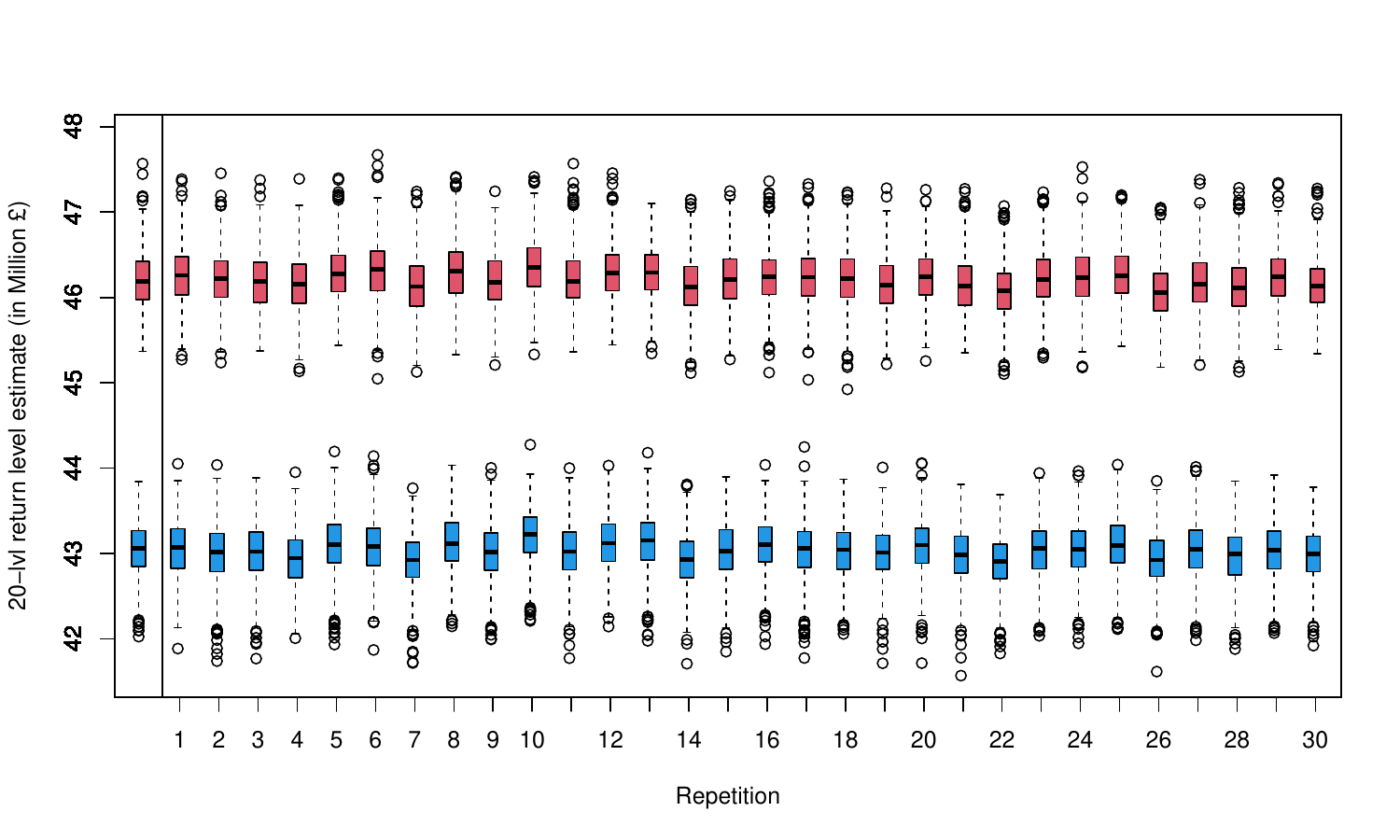}
\includegraphics[scale=0.6]{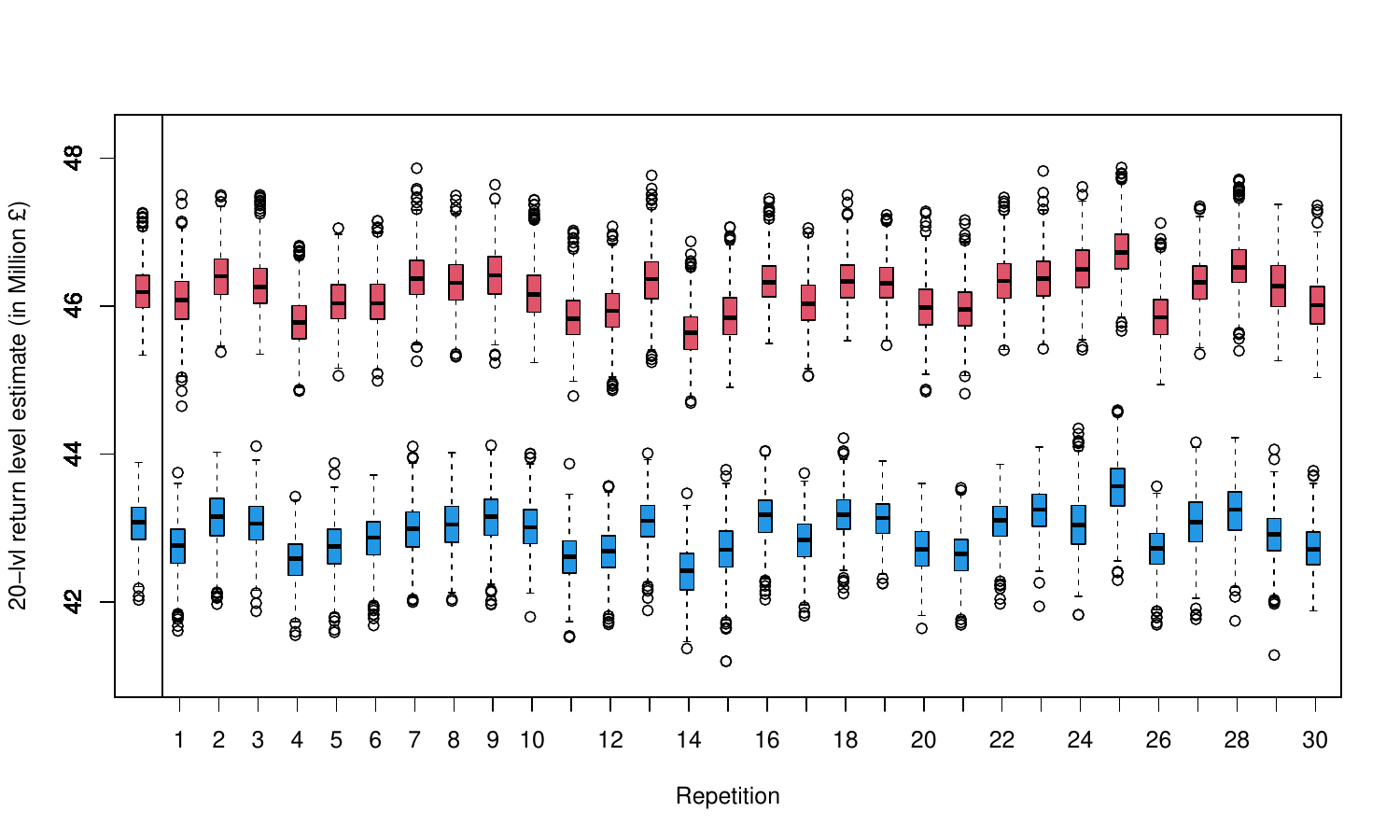}
  \caption{Boxplots of the 1000 simulated 20-year return levels for the portfolio scenarios P3 (top panel) and P4 (lower panel). The red/blue boxplots show the estimates based on the upper/lower conservative bound. The first 30 boxplots of 100 replicates are shown. The two leftmost are boxplots of return level estimates from our original portfolio P0 for comparison.
\label{fig.boxplots20}
}
\end{center}
\end{figure}

Table~\ref{table.sensitivity20} and Figure~\ref{fig.boxplots20} are the corresponding table and boxplots for the 20-year return level in our sensitivity study. We again see that the additional uncertainty in the damage ratio makes little difference to the estimated quantiles in Table~\ref{table.sensitivity20}. It is clearly visible in the box plots of Figure~\ref{fig.boxplots20} that the between-simulation variability is relatively small compared to the within-simulation variability. For P3, $\sigma^2_b/\sigma^2_w$ is approximately 1/200, for both the upper and lower bounds, whereas for P4 the ratio is approximately 1/11. 

\begin{table}[h!]
  \begin{center}
\begin{tabular}{c||c|c|c|c}
  Scenario & Lower 2.5\% quantile & Lower median & Upper median & Upper 97.5\% quantile\\
  \hline
  P0 & 42.42 & 43.07 & 46.22 & 46.88 \\
  P1 & 44.60 & 45.24 & 48.49 & 49.14 \\
  P2 & 40.30 & 40.89 & 43.92 & 44.59 \\
  P3 & 42.37 & 43.05 & 46.22 & 46.90 \\
  P4 & 42.00 & 42.96 & 46.21 & 47.20
  \end{tabular}
\caption{\label{table.sensitivity20} Statistics for the 20-year return level. Estimated 2.5\% and 50\% quantiles from the conservative lower bounds and the 50\% and 97.5\% quantiles from the conservative upper bounds in million £s to 4s.f. for different portfolio scenarios. For scenarios P3 and P4 the estimates are from a single pooled sample of all of the $R\times M$ simulations of the return level.}
  \end{center}
  \end{table}

\section{Alternative approaches}
\label{sec.alternativeApproaches}
If upper and lower concentration inequalities were available for the order statistics of the yearly losses for the collection of 1000 or 10000 years outputted by the flood-event model, rather than for the individual losses themselves, then bounds on the quantiles for the relevant losses could be obtained directly from the inequalities without any simulation. For example, concentration inequalities on the 100th highest loss of a total of 1000 years outputted from the flood-event model would provide upper and lower bounds on the point estimate and quantiles of the 10-year return level. The only work of which we are aware of in this area is \cite{BoucheronThomas2012}, which derives concentration inequalities for independent and identically distributed random variables whose cumulative distribution function has a known functional form and, for most of the inequalities, the random variables can be represented by a density with a non-increasing hazard. None of these conditions holds in our case, and any generalisation to include our scenario would require substantial work if it were even possible.

In some scenarios \cite[e.g.,][and references therein]{ScottMetzler2015} it is possible to apply importance sampling to the generation of ``events'', so that fewer non-extreme events and more extreme events are generated. Each generated event is then given a weight: the ratio of the density of the outcome under the generative model that is assumed to be a truth and under the model that favours more extreme years. In our situation the model that generates a single flood event over the region of space covered by an entire portfolio is very complex, involving simulation of a particularly extreme location to be conditioned on and then back-transformation via a different empirical cumulative distribution function at each of thousands of river gauge locations spanning the portfolio region \cite[][]{Lambetal2010,KeefTawnLamb2013,ToweTawnLambSherlock2019}. This output is then spatially interpolated to the locations of all the risks in the portfolio; for each risk, the interpolated level is combined with information about the risk (such as elevation) to give a probability of flooding and a beta distribution for the fraction of the insured value lost if a flood occurs; simulating from this distribution for each subrisk at the risk and summing gives the loss at that risk, and summing over all risks gives the total loss for the event. The number of events in a year is usually simulated from a Poisson distribution; summing over all events in a year gives the loss for that year. The true density/mass function for the loss distributions for a given year is, therefore, not tractable. Even if it were tractable, it would be far from straightforward to create a suitable proposal distribution for the joint distribution across all risks of the flood probabilities and beta parameters for each risk. Moreover, importance sampling is typically used for a particularly high quantile, such as the $0.999$th quantile. The proposed events and years worth of events would then typically be unsuitable for estimating lower quantiles. We are interested in return levels from the two-year ($0.5$ quantile) and up, so a careful balance between the gain in efficiency for high quantiles and the loss of efficiency for low quantiles would need to be maintained.

\end{document}